\documentclass[traditabstract,twocolumns]{aa}

\usepackage{txfonts}
\usepackage{graphicx}	
\usepackage{amsmath}	
\usepackage{amssymb}	
\usepackage[table,x11names]{xcolor}     
\usepackage{soul}       
\usepackage{lineno}    
\usepackage{array}
\usepackage[colorlinks=true,linkcolor=blue,citecolor=blue, urlcolor=blue]{hyperref}
\usepackage{adjustbox} 
\bibpunct{(}{)}{;}{a}{}{,}

\usepackage{newtxtext,newtxmath}

\usepackage{orcidlink}

\makeatletter
\newcommand{\mycustombox}[1]{%
  \fbox{%
    \begin{minipage}{\linewidth-2\fboxsep-2\fboxrule}
      \strut \textbf{Prepared for submission} --- \textit{Physical Review D}\hfill Draft version: \today\strut
    \end{minipage}%
  }%
}
\patchcmd{\aa@maketitle}{\fbox}{\mycustombox}{}{}
\makeatother

\makeatletter
\newcommand{\customcite}[2]{\hyper@@link[cite]{}{cite.#1}{#2}}
\renewcommand*\aa@textidlineempty{\aa@headings}
\renewcommand*\aa@headfont{\small\mathversion{normal}}
\makeatother


\defcitealias{Millon2020_TDC1}{Paper\,I}
\defcitealias{Birrer2020_TDC4}{Paper\,IV}

\newcommand{\lcdm}{$\mathrm{\Lambda CDM}$\xspace}
\newcommand{\lenstro}{\textsc{lenstronomy}\xspace}
\newcommand{\glee}{\textsc{glee}\xspace}
\newcommand{\squirrel}{\textsc{squirrel}\xspace}
\newcommand{\ppxf}{\textsc{ppxf}\xspace}
\newcommand{\Hc}{\ensuremath{H_0}\xspace}
\newcommand{\Om}{\ensuremath{\Omega_{\rm m}}\xspace}
\newcommand{\Ddt}{\ensuremath{D_{\Delta t}}\xspace}
\newcommand{\Ds}{\ensuremath{D_{\rm s}}\xspace}
\newcommand{\Dd}{\ensuremath{D_{\rm d}}\xspace}
\newcommand{\Dds}{\ensuremath{D_{\rm ds}}\xspace}

\newcommand{\zs}{\ensuremath{z_{\rm s}}\xspace}
\newcommand{\zd}{\ensuremath{z_{\rm d}}\xspace}

\newcommand{\reff}{\ensuremath{R_{\rm eff}}\xspace}
\newcommand{\thetaE}{\ensuremath{\theta_\mathrm{E}}\xspace}

\newcommand{\imagepos}{\ensuremath{\boldsymbol{\theta}}\xspace}

\newcommand{\timedelay}{\ensuremath{\Delta t\xspace}}
\newcommand{\timedelayat}[2]{\ensuremath{\timedelay_{\rm #1 #2}}\xspace}
\newcommand{\lint}{\ensuremath{\lambda_{\rm int}}\xspace}
\newcommand{\kext}{\ensuremath{\kappa_{\rm ext}}\xspace}
\newcommand{\nmax}{\ensuremath{n_{\rm max}}\xspace}

\newcommand{\flexshift}{\ensuremath{\Delta_3x}\xspace}

\newcommand{\fifteenthirtyseven}{J1537$-$3010\xspace}

\newcommand{\ir}{\ensuremath{\mathrm{F160W}}\xspace}
\newcommand{\uvisone}{\ensuremath{\mathrm{F814W}}\xspace}
\newcommand{\uvistwo}{\ensuremath{\mathrm{F475X}}\xspace}

\newcommand{\cmark}{$\checkmark$}
\newcommand{\xmark}{$\times$}

\def\ks{${\, \mathrm{km}\, \mathrm{s}^{-1}}$\xspace}
\newcommand{\arcs}{\ensuremath{^{\prime \prime}}}

\begin{document}


\title{{\Large TDCOSMO XXVIII.}\\The Hubble constant from the quadruply lensed quasar \fifteenthirtyseven with precise time delays}

\author{
A.~Galan\inst{\ref{unige}}\corrauth{aymeric.galan@gmail.com}\orcidlink{0000-0003-2547-9815}
\and
A.~G.~Schweinfurth \inst{\ref{tum},\ref{mpa}}\corrauth{allansch@mpa-garching.mpg.de}\orcidlink{0000-0002-8274-7196}
\and
S.~H.~Suyu\inst{\ref{tum},\ref{mpa}}\orcidlink{0000-0001-5568-6052}
\and
D.~P.~Johnson \inst{\ref{lupm}}\orcidlink{0000-0002-0311-2513}
\and
A.~Chawla \inst{\ref{uliege}, \ref{Armargh}}\orcidlink{0009-0000-9327-6695}
\and
D.~Sluse \inst{\ref{uliege}}\orcidlink{0000-0001-6116-2095}
\and
G.~V.~Kharchilava \inst{\ref{fermi}}
\and
E.~Buckley-Geer \inst{\ref{fermi}}
\and
H.~Lin \inst{\ref{fermi}}
\and
S.~Knabel \inst{\ref{ucla}}\orcidlink{0000-0001-5110-6241}
\and
W.~Sheu \inst{\ref{ucla}}\orcidlink{0000-0003-1889-0227}
\and
S.~Ertl \inst{\ref{tum},\ref{mpa}}
\and
T.~Anguita \inst{\ref{unab}}\orcidlink{0000-0003-0930-5815}
\and
S.~Birrer \inst{\ref{stonybrook}}\orcidlink{0000-0003-3195-5507}
\and
F.~Courbin \inst{\ref{barca1},\ref{barca2},\ref{barca3}}\orcidlink{0000-0003-0758-6510}
\and
F.~Dux \inst{\ref{epfl},\ref{santiago}}\orcidlink{0000-0003-3358-4834}
\and
M.~Millon \inst{\ref{unigeth}}\orcidlink{0000-0001-7051-497X}
\and
V.~Motta \inst{\ref{valpa}}\orcidlink{0000-0003-4446-7465}
\and
S.~Schuldt \inst{\ref{finca},\ref{helsinki},\ref{inaf}}
\and
T.~Treu\inst{\ref{ucla}}\orcidlink{0000-0002-8460-0390}
\and
D.~M.~Williams \inst{\ref{ucla}}\orcidlink{0000-0002-8386-0051}
\and
M.~Cappellari \inst{\ref{oxford}}\orcidlink{0000-0002-1283-8420}
\and
K.~C.~Wong\inst{\ref{utokyo}}\orcidlink{0000-0002-8459-7793}
\and
D.~Eckert\inst{\ref{unige}}\orcidlink{0000-0001-7917-3892}
}

\institute{
Department of Astronomy, University of Geneva, Chemin d’Ecogia 16, 1290 Versoix, Switzerland
\label{unige}
\and
Technical University of Munich, TUM School of Natural Sciences, Department of Physics, James-Franck-Str. 1, 85748 Garching, Germany
\label{tum}
\and
Max-Planck-Institut für Astrophysik, Karl-Schwarzschild-Str. 1, 85748 Garching, Germany
\label{mpa}
\and
Laboratoire Univers et Particules de Montpellier (LUPM), CNRS \& Université Montpellier (UMR-5299), Parvis Alexander Grothendieck, F-34095 Montpellier Cedex 05, France
\label{lupm}
\and 
STAR Institute, University of Li{\`e}ge, Quartier Agora, All\'ee du six Ao\^ut 19c, 4000 Li\`ege, Belgium \label{uliege}
\and 
Armagh Observatory and Planetarium, College Hill, Armagh BT61 9DG, N. Ireland, UK \label{Armargh}
\and
Fermi National Accelerator Laboratory, P. O. Box 500, Batavia, IL 60510, USA
\label{fermi}
\and
Department of Physics and Astronomy, University of California, Los Angeles, CA 90095, USA
\label{ucla}
\and
Instituto de Astrof\'isica, Departamento de F\'isica y Astronom\'ia, Universidad Andres Bello, Chile
\label{unab}
\and
Department of Physics and Astronomy, Stony Brook University, Stony Brook, NY 11794, USA
\label{stonybrook}
\and
Institut de Ciències del Cosmos (ICCUB), Universitat de Barcelona (IEEC-UB), Martí i Franquès 1, 08028 Barcelona, Spain
\label{barca1}
\and
75 Institució Catalana de Recerca i Estudis Avançats (ICREA), Passeig de Lluís Companys 23, 08010 Barcelona, Spain
\label{barca2}
\and
76 Institut de Ciencies de l’Espai (IEEC-CSIC), Campus UAB, Carrer de Can Magrans, s/n Cerdanyola del Vallés, 08193 Barcelona, Spain
\label{barca3}
\and
Laboratory of Astrophysics, Institute of Physics, École Polytechnique Fédérale de Lausanne (EPFL), Observatoire de Sauverny, 1290 Versoix, Switzerland
\label{epfl}
\and
European Southern Observatory, Alonso de Córdova 3107, Vitacura, Santiago, Chile
\label{santiago}
\and
Département de Physique Théorique, Université de Genève, 24 quai Ernest-Ansermet, CH-1211 Genève 4, Switzerland
\label{unigeth}
\and
Finnish Centre for Astronomy with ESO (FINCA), University of Turku, FI-20014 Turku, Finland
\label{finca}
\and
Department of Physics, P.O. Box 64, University of Helsinki, FI-00014 Helsinki, Finland
\label{helsinki}
\and
INAF - IASF Milano, via A. Corti 12, I-20133 Milano, Italy
\label{inaf}
\and
Instituto de F\'{\i}sica y Astronom\'{\i}a, Universidad de Valpara\'{\i}so, Av. Gran Breta\~na 1111 Playa Ancha, Valpara\'{\i}so, Chile
\label{valpa}
\and
Sub-Department of Astrophysics, Department of Physics, University of Oxford, Denys Wilkinson Building, Keble Road, Oxford, OX1 3RH, UK
\label{oxford}
\and
Research Center for the Early Universe, Graduate School of Science, The University of Tokyo, 7-3-1 Hongo, Bunkyo-ku, Tokyo 113-0033, Japan
\label{utokyo}
}

\abstract{
We present the first measurement of the Hubble constant (\Hc) from the quadruply-lensed quasar \fifteenthirtyseven, which has optically-measured time delays at $\sim2 \%$ precision. We combine these delays with multi-band imaging data from the Hubble Space Telescope (HST) and model the system with two independent software and teams. We adopt a mass profile that is maximally degenerate with \Hc to fully incorporate the mass-sheet degeneracy in the error budget, with nuisance parameters constrained by spatially resolved stellar kinematics from the Multi Unit Spectroscopic Explorer (MUSE) and a line-of-sight (LoS) analysis using the Euclid Flagship simulation. The entire analysis is performed blindly to \Hc, distances and mass density slope of the main deflector. After unblinding, we measure $\Hc = 75.5^{+9.3}_{-5.8}\ {\rm km\,s^{-1}\,Mpc^{-1}}$, corresponding to a $10\%$ precision measurement from a single system. This precision is driven by conservative lens modeling assumptions including differences between lens modeling methods and the limited stellar kinematics constraints, from which we infer a total mass-sheet parameter $\lambda\equiv(1-\kext)\lint = 0.89^{+0.11}_{-0.06}$ ($\lint=0.89^{+0.09}_{-0.07}$) that is consistent with current results for elliptical galaxies ($\lambda \approx 1$). Our lensing constraints, stellar kinematic measurements and LoS characterization will be included in subsequent population-level measurements of \Hc. Moreover, our lens models combined with future near-infrared spectroscopy and imaging from the James Webb Space Telescope will further reduce the uncertainties on \Hc from \fifteenthirtyseven alone, bringing it closer to the few percents precision of the time delays.
}


\titlerunning{Direct \Hc measurement from \fifteenthirtyseven}
\authorrunning{A.~Galan, A.~G.~Schweinfurth et al.}

\maketitle
\nolinenumbers 



\section{Introduction}

\begin{figure*}[!ht]
    \centering
    \includegraphics[width=\linewidth]{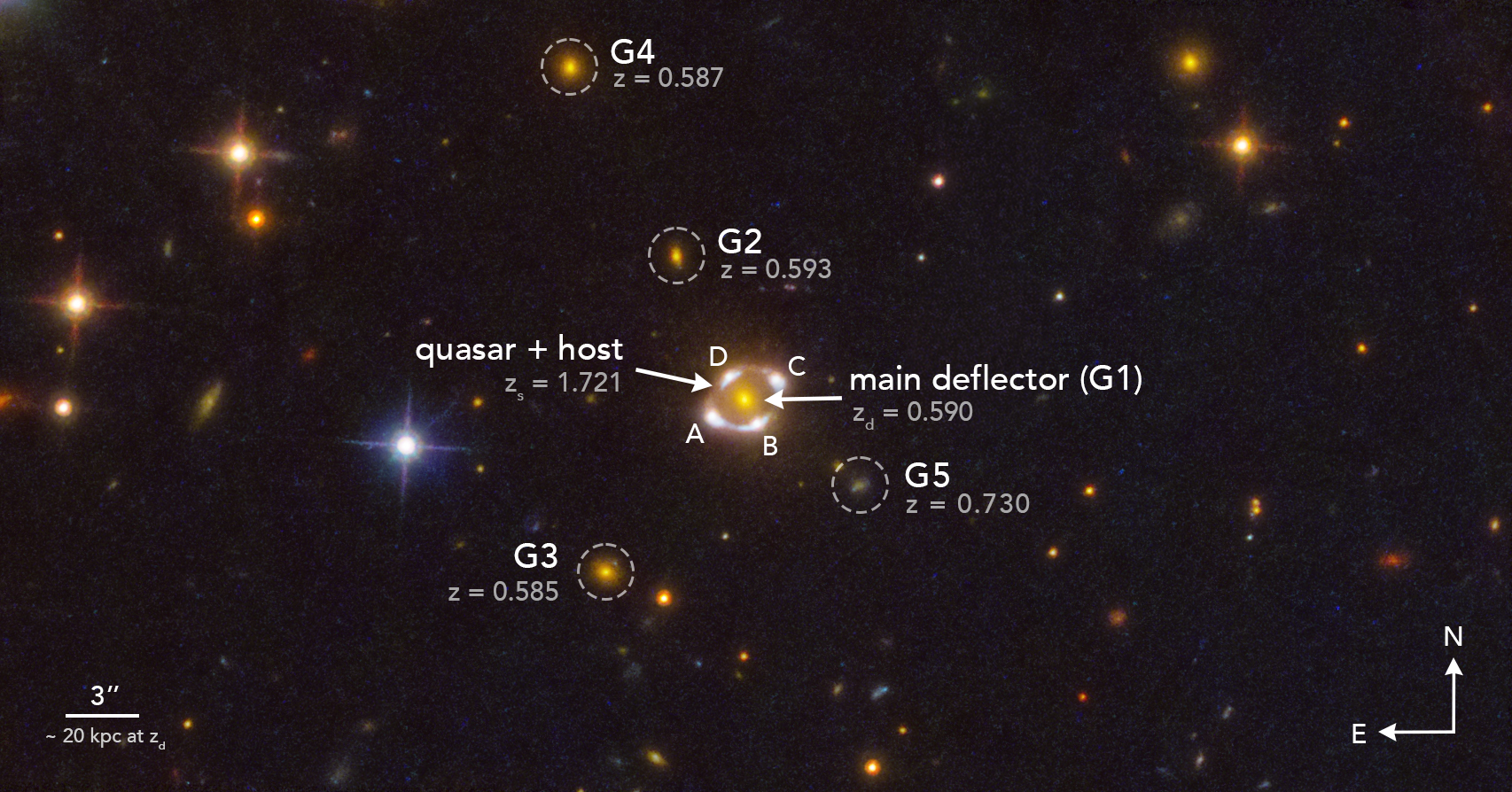}
    \caption{Color composite image of the quadruply lensed quasar \fifteenthirtyseven built from HST/WFC3 data (red: \ir, green: \uvisone, blue: \uvistwo). Labels for the quasar images (A to D), main deflector (G1) and identified perturbers (G2 to G5) are also indicated, as well as the angular scale and orientation of the image. Additional credits to Tian Li for the color image.}
    \label{fig:color_image}
\end{figure*}

Time-delay cosmography \citep{Refsdal1964} exploits the arrival time difference between photons emitted by strongly lensed sources to measure the Hubble constant \citep[\Hc; for recent reviews, see e.g.,][]{,2024SSRv..220...48B,2024hct..book..251T}. This method is fully independent from others such as those based on the cosmic microwave background \citep[e.g.,][]{2020A&A...641A...6P,2023PhRvD.108b3510B} or on cosmic distance ladders \citep[e.g.,][]{2022ApJ...934L...7R,2025ApJ...985..203F}. Therefore, time-delay cosmography plays a key role in the context of the Hubble tension, where unknown systematic biases and profound theoretical changes are still two possible outcomes \citep[for a recent review of cosmological tensions, see][]{2025PDU....4901965D}.

Time-delay cosmography applies to the local Universe, as it measures ratios of absolute distances typically at redshifts $z \sim 0.2$ to $z \sim 2$; it is also direct as it measures these distances without relying on any calibration steps. The Time Delay COSMOgraphy (TDCOSMO) collaboration is leading the effort to reaching percent-level precision on \Hc using this method. Our latest cosmological results, $\Hc = 71.6^{+3.9}_{-3.3}\ {\rm km\,s^{-1}\,Mpc^{-1}}$ in flat $\Lambda$ Cold Dark Matter (\lcdm) cosmology, were presented in \citet{2025A&A...704A..63T} based on a hierarchical analysis combining 8 time-delay lensed quasars and 15 galaxy-galaxy strong lenses with high-quality lensing and stellar kinematics constraints.

Here we add a new lensed quasar \fifteenthirtyseven at RA = $234.3556$, Dec = $-30.1713$ to the sample, shown in Fig.~\ref{fig:color_image} using our Hubble Space Telescope (HST) observations. We present the end-to-end analysis of this quadruply imaged lensed quasar, the measurement of \Hc from this system only, as well as the constraints that will be included in the next TDCOSMO milestone measurements of \Hc. This prominent system has been discovered independently by \cite{Lemon2019} and \cite{Delchambre2019}, and spectroscopically confirmed in \citet{2021ApJ...921...42S}. Both teams used the Gaia data release 2 \citep[GDR2;][]{2018A&A...616A...1G} to search for lenses, although with different methods. \cite{Lemon2019} used a combination of multiple datasets (Gaia detections, Gaia astrometry, PanSTARRS, unWISE) to build a lens-candidate catalog. \cite{Delchambre2019} used supervised machine-learning techniques to look for lenses in GDR2. Both teams spectroscopically confirmed the lensed quasar nature of \fifteenthirtyseven.

\begin{figure*}[!ht]
    \centering
    \includegraphics[width=\linewidth]{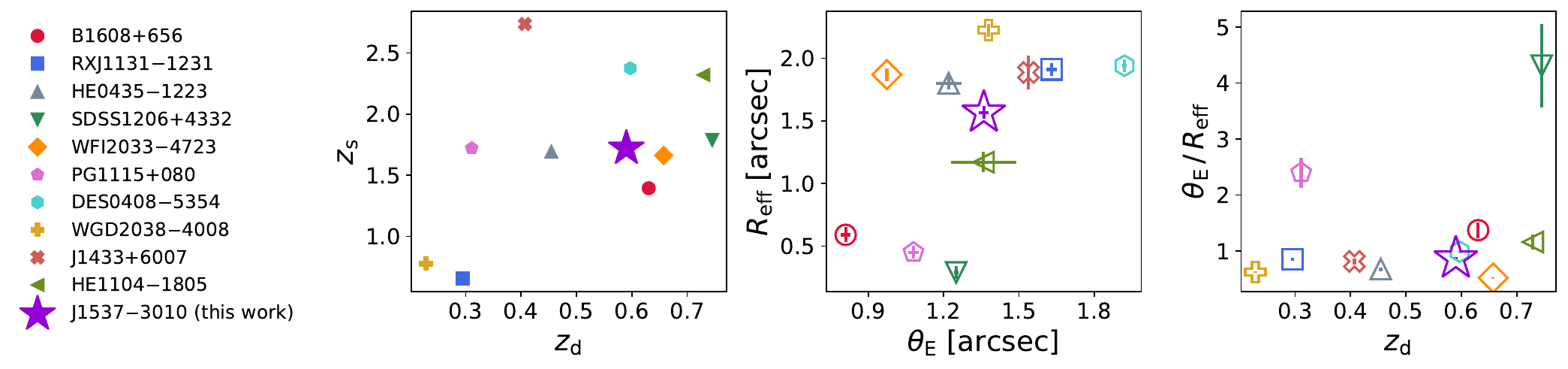}
    \caption{The place of \fifteenthirtyseven within the current TDCOSMO sample of lensed quasars: deflector redshift \zd, source redshift \zs, Einstein radius \thetaE and effective (half-light) radius \reff of the main deflector. For \thetaE and \reff, 1-$\sigma$ error bars are also indicated. Values for other systems were taken from \citet{2025A&A...704A..63T,2026A&A...706A.270P,2026arXiv260414145S}, Williams et al. (in prep.). The \reff value for ${\rm J}1433\!+\!6007$ has been remeasured on a larger cutout and thus slightly differs from \citet{2026arXiv260414145S}.}
    \label{fig:tdcosmo_sample}
\end{figure*}

The source (quasar + host) redshift $z=1.721$ has been measured from spectra taken with the William Herschel Telescope \citep{Lemon2019} \footnote{\citet{Lemon2019} originally quoted \zs two decimals ($1.72$), while \citet{2021ApJ...921...42S} added a third decimal ($1.721$).}. A preliminary deflector redshift of $z=0.592$ based on spectroscopic data obtained with the Inamori Magellan Areal Camera and Spectrograph (IMACS) was reported in \citet{Schmidt2023}. However, we measure again the redshifts for the purpose of this analysis to find $\zd=0.590$ and $\zs=1.721$ for the lens and source, respectively (Sect.~\ref{ssec:redshifts}). We summarize global properties of \fifteenthirtyseven and compare those to the current TDCOSMO sample in Fig.~\ref{fig:tdcosmo_sample}. This system falls in the mid-range of the sample in terms of source redshift and Einstein radius, and the higher end of the lens redshift and effective radius distribution.

Although not used in our cosmographic measurements, when necessary we assume a fiducial Lambda cold dark matter (\lcdm) cosmology with $\Hc = 70\ {\rm km}\,{\rm s}^{-1}\,{\rm Mpc}^{-1}$ and $\Omega_{\rm m} = 0.3$. This cosmology leads to angular sizes of $6.6\ {\rm kpc\, arcsec}^{-1}$ at the lens redshift \zd and $8.5\ {\rm kpc\, arcsec}^{-1}$ at the source redshift \zs. We use the shortened notations $D_i \equiv D_{\rm A}(0,\,z_i)$ and $D_{ij} \equiv D_{\rm A}(z_i,\,z_j)$ for angular diameter distances.

The paper is organized as follows. In Sect.~\ref{sec:tdcosmo_method} we briefly recall the main principles of time-delay cosmography and how we can mitigate the main degeneracy affecting \Hc measurements. In Sect.~\ref{sec:data_sets} we present the main data sets used in our analysis. We summarize the time-delay measurement in Sect.~\ref{sec:time_delays}. We present the first measurements of spatially resolved stellar kinematics in Sect.~\ref{sec:kinem_measurements}, as well as the environment and line-of-sight (LoS) analysis of the system in Sect.~\ref{sec:environment_and_los}. The constraints from lens modeling are described in Sect.~\ref{sec:lens_modeling}, which we combine with the constraints from stellar kinematics and LoS analysis in Sect.~\ref{sec:dynamical_weighting}. The final measurement of \Hc is described in Sect.~\ref{sec:cosmo_measurements} and we conclude this work in Sect.~\ref{sec:conclusion}.

To prevent experimenter bias, our analysis is performed blindly with respect to key quantities that may be informative about the final inferred \Hc value. In addition to \Hc, the blinded quantities include the time-delay distance, and angular diameter distances and the mass density slope of the main deflector. In practice, this is achieved by removing the mean or median of these quantities, before producing texts and figures. Once the code, results, and manuscripts (with appropriate placeholders in texts and figures) have been internally reviewed, the analysis is frozen and no major modifications can be made. Following our internal publication workflow, this analysis was reviewed by the collaboration prior to unblinding the \Hc measurement, including two designated reviewers (E.P., S.B.), and three code reviewers (E.P., K.C.W., W.S.). After unblinding\footnote{Unblinding took place on August 7, 2026, 5am CEST.}, figures and tables with placeholder values were updated.

\section{Time-delay cosmography \label{sec:tdcosmo_method}}

\subsection{Theory}

In gravitational lensing, the lens equation 
maps a source position $\boldsymbol{\beta}$ in the source plane to observed image 
positions $\boldsymbol{\theta}$ in the lens plane via
\begin{equation}
\label{eq:lens_equation}
    \boldsymbol{\beta} = \boldsymbol{\theta} - \boldsymbol{\alpha}(\boldsymbol{\theta}) \ ,
\end{equation}
where $\boldsymbol{\alpha}(\boldsymbol{\theta})$ is the scaled deflection angle.

The observed time delay between a pair of lensed images A and B is
\begin{equation}
    \label{eq:def_time_delays}
    \Delta t_{\rm AB} = \frac{D_{\Delta t}}{c}\,\Delta\phi_{\rm AB} \ ,
\end{equation}
where $\Delta\phi_{\rm AB} \equiv \phi(\boldsymbol{\theta}_{\rm A}) - \phi(\boldsymbol{\theta}_{\rm B})$ 
is the difference in the Fermat potential between the two images. In the case of a single lens plane, the Fermat potential is
\begin{equation}
    \phi(\boldsymbol{\theta}) = \frac{(\boldsymbol{\theta} - \boldsymbol{\beta})^2}{2} 
    - \psi(\boldsymbol{\theta}) \ ,
\end{equation}
where $\psi(\boldsymbol{\theta})$ is the lensing potential. In this single-plane setting, the 
time-delay distance is
\begin{equation}
\label{eq:def_ddt}
    D_{\Delta t} = (1+z_{\rm d})\frac{\Dd\,D_{\rm s}}{D_{\rm ds}} \ ,
\end{equation}
where $\Dd$, $D_{\rm s}$, and $D_{\rm ds}$ are the angular diameter distances 
from the observer to the deflector, from the observer to the source, and from the 
deflector to the source, respectively, and $z_{\rm d}$ is the deflector redshift. Cosmological parameters, in particular \Hc, are encapsulated in the time-delay distance \Ddt. In the context of \lcdm, angular diameter distances are inversely proportional to \Hc, hence $\Hc \propto D_{\Delta t}^{-1}$ from Eq.~\ref{eq:def_ddt}. In contrast, $D_{\Delta t}$ only weakly depends on other cosmological parameters (e.g., $\Omega_{\rm m}$).

In a multi-plane setting, in which additional deflector planes are included explicitly in the models, Eq.~\ref{eq:lens_equation} must be evaluated iteratively \citep[][]{1992grle.book.....S}. The Fermat potential $\Delta\phi_{\rm AB}$ then receives contributions from each deflector plane, and the time-delay distance generalizes to an effective time-delay distance $\Ddt$. However, to avoid clutter, we simply employ the \Ddt notation throughout the paper for both our single-plane and multi-plane models, and refer the reader to \citet{Shajib2020} for the detailed mathematical formalism.

\subsection{Mass-sheet degeneracy \label{ssec:mst_theory}}

The mass-sheet degeneracy \citep[MSD;][]{Falco1985} is the dominant source of 
systematic uncertainty when modeling images of strong lenses, thus it is a significant source of systematic uncertainty on \Hc when measured with time-delay cosmography. The MSD arises from the mass-sheet transformation (MST) that is a rescaling of the convergence $\kappa(\boldsymbol{\theta}) \equiv \nabla^2 \psi(\boldsymbol{\theta})/2$ as
\begin{equation}
\label{eq:mst}
    \kappa \to \kappa_\lambda(\boldsymbol{\theta}) = \lambda\,\kappa(\boldsymbol{\theta}) 
    + (1 - \lambda) 
\end{equation}
combined with a rescaling of the (unobserved) source coordinates $\boldsymbol{\beta} \to \boldsymbol{\beta}_\lambda = \lambda\boldsymbol{\beta}$. Mathematically, the MST parameter $\lambda$ corresponds to an infinite ``sheet'' of mass. The MST leaves all lensing observables invariant---image
positions and flux ratios---except the time delays. As such, the time-delay distance (equivalently, the Fermat potential, see Eq.~\ref{eq:def_time_delays}) is rescaled $\Ddt \to \lambda\,\Ddt$ by the MST.

It is particularly useful to decompose $\lambda$ into contributions that are internal or external to the main deflector, as
\begin{align}
\label{eq:mst_decomp}
    \lambda \equiv (1-\kext)\,\lambda_\mathrm{int} \ ,
\end{align}
where $\lambda_\mathrm{int}$ describes a transformation of the main deflector's mass distribution, and $\kappa_\mathrm{ext}$ is the convergence arising from matter along the line of sight \citep[e.g.][]{Koopmans2003,Saha_2006,Schneider_2013,Birrer_2016,Birrer_2020}. $\kext$ can in turn be written as 
\begin{align}
\label{eq:los_breakdown}
    1-\kext = \frac{(1-\kappa_\mathrm{d})(1-\kappa_\mathrm{s})}{1-\kappa_\mathrm{ds}} \ ,
\end{align}
where $\kappa_\mathrm{d}$, $\kappa_\mathrm{s}$ and $\kappa_\mathrm{ds}$ are the weak lensing convergence terms between the observer and lens, observer and source, and deflector and source, respectively, in the absence of the main deflector's convergence \citep{Birrer_2020,Fleury_2021}.

The MSD can be efficiently broken through the combination of (1) simple physical arguments that discard unphysical mass distributions, (2) the careful characterization of the LoS contribution to the net lensing effect, and (3) measurements of stellar kinematics of the main deflector. In particular, stellar kinematics probe the (three-dimensional) gravitational potential of the deflector, whereas lensing is sensitive to the full potential projected along the LoS. Beyond their ability to break the internal MSD, stellar kinematics also provide a measurement of the angular diameter distance to the deflector, $\Dd$, through the projected Jeans equation \citep[e.g.,][]{2009A&A...507L..49P,2015JCAP...11..033J}. Therefore, $\Ddt$ and $\Dd$ jointly provide constraints on cosmological parameters, ultimately improving both the accuracy and the precision of \Hc \citep[e.g.,][]{Jee+2019}. Single-aperture velocity dispersions are limited by the mass-anisotropy degeneracy \citep[MAD;][]{1982MNRAS.200..361B}, but spatially resolved kinematics additionally constrain the radial slope of the mass profile, simultaneously mitigating both the internal MSD and the MAD \citep[e.g.,][]{STA18,Yildirim+2023, 2023A&A...673A...9S, 2025A&A...704A..63T, 2025A&A...701A.280W}.

\section{Main data sets \label{sec:data_sets}}

In this section, we report the data sets we obtained specifically for this analysis. Other data sets we use in this work include, for instance, public Gaia \citep{2018A&A...616A...1G} or the DECam Local Volume Exploration Survey \citep[DELVE;][Drlica-Wagner et al., in prep.]{2021ApJS..256....2D} data releases.

\subsection{Ground-based monitoring campaign \label{ssec:monitoring}}

\begin{figure}[!t]
    \centering
    \includegraphics[width=\linewidth]{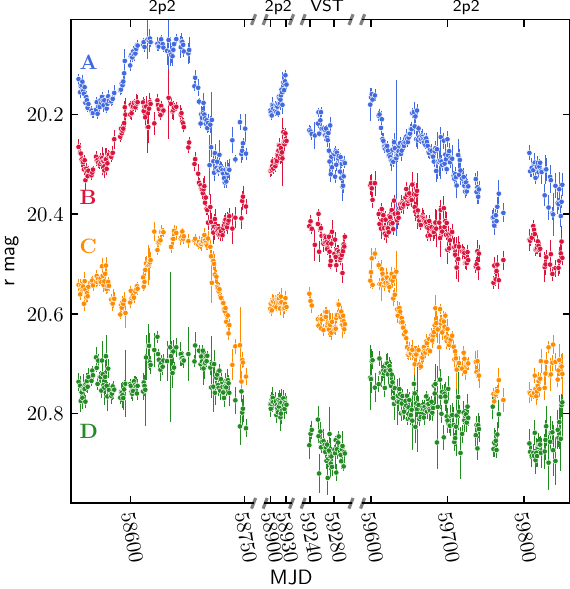}
    \caption{Light curves in $r$ band extracted from the four-season long monitoring campaign (Sect.~\ref{ssec:monitoring}), for each quasar image (see Fig.~\ref{fig:color_image} for the image labeling). Figure reproduced from \citet{2025A&A...697A.139D}.}
    \label{fig:light_curves}
\end{figure}

The system \fifteenthirtyseven was monitored for four seasons spanning 2019–2022, split between the MPG/ESO 2.2-meter (2p2) telescope and the VLT Survey Telescope (VST) \citep{2025A&A...697A.139D}. The first full season ran from February to October 2019 at the 2p2, followed by two pandemic-shortened seasons in early 2020 (2p2) and early 2021 (VST), and a final complete season at the 2p2 from January to September 2022. The combined dataset amounts to roughly two and a half effective seasons of coverage, and the light curves were extracted using the photometric pipelines \textsc{lightcurver} \citep{Dux2024} and \textsc{starred} \citep{Millon2024, Michalewicz2023}. The resulting light curves are reproduced in Fig.~\ref{fig:light_curves} from \citet{2025A&A...697A.139D}.

\subsection{High-resolution imaging data}

We primarily use the HST imaging data of \fifteenthirtyseven obtained in Cycle 26 (PI: Treu; PID: 15652), using the Wide Field Camera 3 (WFC3) in the three filters, \ir (IR), \uvisone and \uvistwo (UVIS). The observations took place on June 22, 2019 with effective exposure times of $\sim2200$ sec for \ir, $\sim1430$ sec for \uvisone and $\sim840$ sec for \uvistwo. We use the same drizzled images already presented by \citet{Schmidt2023}, rotated such that the North is up and East is to the left. The final pixel scales are $0\farcs08$ in the \ir, and $0\farcs04$ in \uvisone and \uvistwo. In Fig.~\ref{fig:color_image} we show a composite color image from these three filters. 

We note that \fifteenthirtyseven has also been observed with the James Webb Space Telescope (JWST) using the Mid Infrared Instrument (MIRI) as part of Cycle 1 (PI: Nierenberg; PID: 2046). These JWST/MIRI data were taken on March 7, 2023, in F560W, F1280W, F1800W and F2100W filters \citep[see also][]{2024MNRAS.535.1652K}. Qualitatively, the F560W data has comparable resolution and signal-to-noise ratio (S/N) of the lensed arc compared to the F160W HST data. However, these JWST/MIRI data were obtained primarily to study the properties of dark matter \citep{2024MNRAS.530.2960N} and hence the choice of filters and instrument were not optimized for time-delay cosmography. Furthermore, in order to avoid the complexity of modeling the MIRI PSF, we do not include these data in the current analysis and leave this task for future work.

\subsection{Spectroscopic data}
\label{ssec:lens_kinematics}

We obtained multi-object slit spectroscopy over a field of view (FoV) of about $5.5 \times 5.5$ arcmin, using the Gemini Multi-Object Spectrograph (GMOS-S) on the Gemini South telescope from Mar.\ to Jun.\ 2022 (PI: Buckley-Geer, program ID GS-2022A-Q-135). Galaxy targets with $i < 23$ were selected using GMOS-S imaging data obtained in the same program. Four multi-object spectroscopy (MOS) masks were designed, and each mask was observed for 2988 seconds ($=$ 747 seconds $\times$ 4 exposures) in each of the blue B600 and red R400 GMOS-S gratings, resulting in a combined spectral coverage of about 4000 to 10000~\AA. Data reduction and redshift measurements were performed using a procedure similar to that described by\cite{BuckleyGeer2020}, which we summarize here in Sect.~\ref{app:ssec:catalog_making}. A total of 60 secure galaxy redshifts were obtained.

We also obtained integral field unit (IFU) spectroscopic observations using the Multi Unit Spectroscopic Explorer (MUSE) instrument on the Very Large Telescope (VLT), together with the GALACSI adaptive Optics system that enables seeing-improved observations \citep{Arsenault2008, Bacon2010, Strobele2012}. The data were obtained in wide field mode (WF) between Feb. and Aug. 2021 (PI: Sluse; PID: 105.20KM.002). The spatial pixel scale is $0\farcs2$ and the mean seeing throughout the observed wavelength range is $\sim 0\farcs7$ FWHM. The data reduction follows the standard steps implemented in the MUSE reduction pipeline (v.2.8.5) called within ESO Reflex reduction environment \citep{Freudling2013}, but with some improvements suggested by \cite{2023A&A...670A...4B}. In particular, we include self-calibration and sky subtraction with the Zurich Atmospheric Purge (ZAP) software \citep{Soto2016}. The data set results from the combination of 8$\times$4 individual exposures of 600\,s, with the lens centered on a different quadrant of the data cube for each set of 4 exposures. This observing strategy yields a field of view of $\sim$ 90\arcs$\times$90\arcs centered on the lens, but the maximum exposure time (5.2\,h) is only achieved for the central 30\arcs$\times$30\arcs region. The spectra cover the range 4700-9348.75 \AA\ and the instrumental dispersion is of $\sigma_{\rm inst} = 120$\ks.

\section{Time delays \label{sec:time_delays}}

\begin{table}
\renewcommand{\arraystretch}{1.4}
\centering
\caption{Time delays and covariance matrix for J1537$-$3010.}
\label{tab:td_j1537}
\begin{tabular}{lrrrr}
\hline\hline
Pair & Delay (days) & \multicolumn{3}{c}{Covariance (days$^2$)} \\
     &              & ${\rm A}-{\rm B}$ & ${\rm A}-{\rm C}$ & ${\rm A}-{\rm D}$ \\
\hline
${\rm A}-{\rm B}$ & $-29.2$ & $0.8$ & $0.1$ & $0.4$ \\
${\rm A}-{\rm C}$ & $8.5$   & $0.1$ & $0.5$ & $0.2$ \\
${\rm A}-{\rm D}$ & $-24.7$  & $0.4$ & $0.2$ & $2.2$ \\
\hline
\end{tabular}
\tablefoot{
The labels of the quasar images are indicated in Fig.~\ref{fig:color_image}. A positive delay implies the second image arrives first. For example, ${\rm A}-{\rm B}= -29.2\ {\rm days}$ means that a feature is first seen in image A and arrives 29.2 days later in B.
}
\end{table}

The time-delay measurements from the light curves of Fig.~\ref{fig:light_curves} were published in \citet{2025A&A...697A.139D}, but we provide a brief overview here (see also Fig.~\ref{fig:color_image} for the image labeling convention). These measurements benefited from the quasar's large and fast intrinsic variations, which are well separated in frequency from the microlensing signal, making the alignment of the four light curves robust and largely insensitive to the choice of extrinsic variation model. The most precise individual delay is ${\rm B} - {\rm C} = 37.7 \pm 0.8\ {\rm days}$ (2\% precision), and combining all delays relative to image B yields a best-case combined precision of 1\% in terms of error contribution to \Hc\ --- the tightest constraints obtained across the entire \citet{2025A&A...697A.139D} sample. The full set of delays and the corresponding $6\times6$ covariance matrix is provided in Table A.17 of \citet{2025A&A...697A.139D} and we report in Table~\ref{tab:td_j1537} the three delays and covariance matrix.

In this work, we consider only the subset of three independent delays relative to image A to follow usual conventions. Since we use the associated covariance matrix, the choice of reference image is irrelevant (it differs by a linear transform) and thus does not affect the precision power of the time delays.

\section{Stellar kinematics of the main deflector \label{sec:kinem_measurements}}

\begin{figure*}
    \centering
    \includegraphics[width=\linewidth]{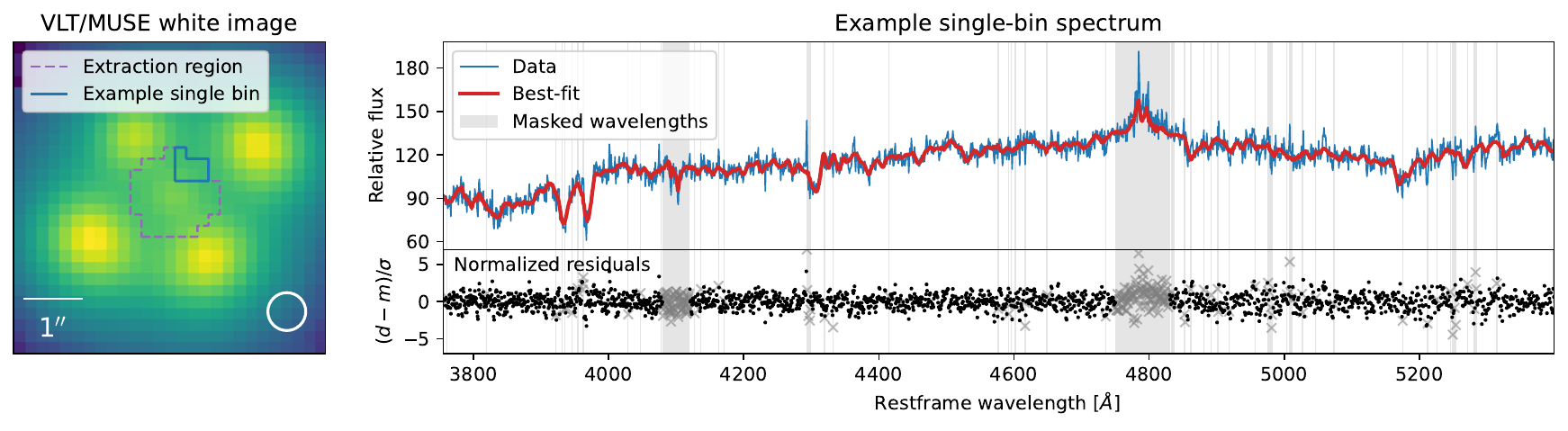}
    \includegraphics[width=0.9\linewidth]{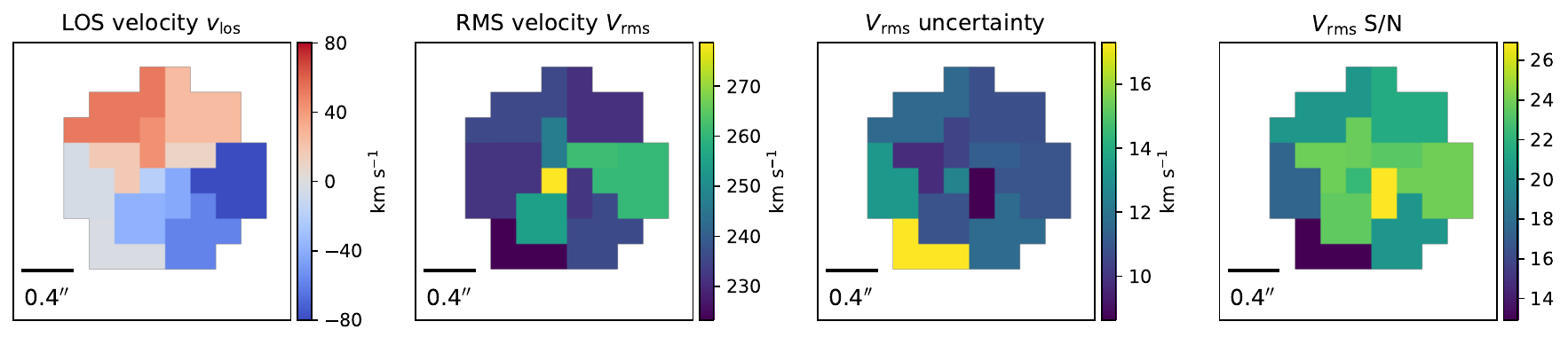}
    \caption{Stellar kinematics measurements for the main deflector using the VLT/MUSE observations. \textit{Top row, left panel}: white image cutout showing the circular region ($0\farcs8$ radius) used for the extraction. The white circle (diameter) shows the mean FWHM of the data over the considered wavelength range. \textit{Top row, right panel}: spectrum from the single spaxel shown in blue on the left panel, along with the best-fit and normalized residuals. The shaded regions are masked out during the fit (quasar and host emission lines and sigma-clipping step). \textit{Bottom row, from left to right}: resulting LoS mean velocity ($v_{\rm los}$), RMS velocity ($V_{\rm rms}$), uncertainty and S/N ($V_{\rm rms}$ divided by the uncertainty).}
    \label{fig:kinem_measur}
\end{figure*}

\subsection{Deflector and quasar redshifts \label{ssec:redshifts}}

We extract resolved stellar kinematics maps of the main deflector using our MUSE data using the Python package \squirrel \footnote{\url{https://github.com/ajshajib/squirrel}} \citep{2026A&A...707A.314S}, which is a pipeline based on Penalized Pixel-Fitting (\ppxf) software\footnote{\url{https://pypi.org/project/ppxf/}} \citep{Cappellari17, Cappellari23}. From an initial spectral fit of the integrated spectrum, we measure $\zd=0.590$ for the main deflector, slightly below literature results. From the quasar spectra we also confirm the source redshift quoted in \citet{2021ApJ...921...42S} $\zs=1.721$. These redshifts are precise up at least up to the last quoted digit.

\subsection{Quasar spectra and PSF modeling \label{ssec:muse_qso_modeling}}

We model the spectra of the quasar in order to separate their flux contribution over the region covered by the main deflector. We used a custom routine that implements $\chi^2$ regression on each wavelength slice of the data cube. A model composed of four Moffat profiles for each quasar image, and of one elliptical S\'ersic (convolved by the Moffat) for the lensing galaxy has been used \citep[for details, see][]{Sluse2017}. Their relative positions were fixed to HST astrometry during the fit. When averaged over the wavelength range used for measuring the stellar kinematics, the full width at half maximum (FWHM) of the Moffat point spread function (PSF) model is ${\rm FWHM}_{\rm Moffat} = 0\farcs67$, with index $\beta_{\rm Moffat} = 1.82$. This model allows us to build a model of the PSF that will be used for the dynamical modeling step (Sect.~\ref{sec:dynamical_weighting}), spectral templates for each quasar image, as well as quasar-subtracted data cube used for systematic checks.

\subsection{Stellar kinematics measurements}

We bin the MUSE spaxels to ensure sufficiently high S/N in each fitted spectrum. We start by defining a circular region centered on the lens with radius of $0\farcs8$, such that we do not perform the extraction in spaxels with significant flux contribution from the quasar and its host. We then bin the spaxels within this restricted region using the \textsc{PowerBin} method \citep{2025MNRAS.544.1432C}, enforcing a minimum specific S/N (sS/N) of 150 \citep[for a definition of sS/N, see][]{2026A&A...707A.314S}. This binning step results in 22 bins. Approximately half of the bins are composed of a single spaxels.

We perform our measurements over the restframe wavelength range 3755-5400~\AA (shown in Fig.~\ref{fig:kinem_measur}), which encompasses all the main absorption features of the deflector while eliminating low S/N regions. Over this wavelength range, we compute the line spread function (LSF) using the model of the median FWHM of MUSE from \citet{2023A&A...670A...4B}. We take the average value between the minimum and maximum wavelengths, leading to 2.6 \AA\ FWHM in the observed frame.

We follow the recipe outlined by \citet{Knabel25} and carefully select stellar template libraries and corrective polynomials. We utilize two commonly-used stellar template libraries, Indo-US \citep{Valdes04_indo_us} and MILES \citep{SanchezBlazquez2006,Falcon-Barroso11}, in their cleaned form as described in \citet{Knabel25}\footnote{The cleaned libraries can be obtained from \url{https://github.com/TDCOSMO/KINEMATICS_METHODS}.}, which are appropriate for the restframe wavelength range of our fits. Indo-US templates are brought to the data resolution in restframe before fitting, while we correct for the MILES resolution by adding the difference in resolution in quadrature to the velocity dispersion after fitting. For all fits, we mask out the restframe wavelength range 4750-4830~\AA\ surrounding the brightest quasar emission lines, as well as 4080-4120~\AA\ surrounding two absorption features likely attributed to the quasar host. We also perform sigma-clipping to mask spectral features that cannot be accurately fitted (see Fig.~\ref{fig:kinem_measur} for the resulting mask). Although we mask out the most prominent quasar features, we include in the fit the quasar spectrum as additional fixed templates (one for each quasar image) obtained from our PSF modeling step in Sect.~\ref{ssec:muse_qso_modeling}, to account for the quasar continuum and subdominant emissions. In Sect.~\ref{app:sec:kinem_qsosub}, we show that performing the extraction on the quasar-subtracted cube leads to very consistent results.

We test a wide range of additive (4-16) and multiplicative (0-3) degrees for polynomial continuum correction. We find that additive degrees necessary to obtain stable measurements lie between 10 and 16, while multiplicative degrees between 0 and 3 are already sufficient. We therefore select additive polynomial degrees \{10, 12, 14, 16\} and multiplicative polynomial degrees \{0, 1, 2, 3\} to include in the set of marginalized models. The result is $4\times4=16$ fits for each of the spatial bins and each template library. We perform a preliminary fit to the spectrum integrated over all spaxels within the circular region, from which we collect all of the templates that contribute at least 1\% of the model to use for the final fits. We do not expect our data to reveal significant stellar population differences between spatial bins \citep[see, e.g.,][]{Knabel24}, and this step helps to ensure that unreasonable templates are not used to fit noisy features or regions of poorly calibrated continuum. We also use this preliminary fit to extract a noise scaling factor, which is the value required to give a reduced $\chi^2 \approx 1$ per binned spectrum. Assuming the spectral fit is good, this factor accounts for over/underestimation of the noise from the data reduction pipeline, which is standard procedure for kinematic fits. Moreover, we identify outlier pixels from the preliminary fit residuals and mask those pixels for the final fits (in addition to the original quasar emission line mask).

\subsection{Bayesian model averaging and covariance matrix}

We marginalize over the choices of libraries and polynomials following the procedure introduced by \citet{Knabel25} using the Bayesian Information Criterion (BIC). For each choice we calculate the weight from the combined $\rm \Delta BIC$ (difference with lowest BIC model) of all binned spaxels. We also calculate the mean, statistical uncertainty, and systematic uncertainty for both the mean velocity $v_{\rm los}$ and velocity dispersion $\sigma_{v,\rm los}$. We construct a covariance matrix whose off-diagonal elements represent the correlated errors between spatial bins and are averaged over the fits of polynomial degrees and template libraries with the average weights from the library test. The off-diagonal terms are all smaller than $0.1\%$. We show in the bottom row of Fig.~\ref{fig:kinem_measur} the extracted $v_{\rm los}$ map and the root-mean-square velocity $V_{\rm rms} = (v_{\rm los}^2 + \sigma_{v,\rm los}^2)^{1/2}$. Since dispersion dominates over the rotational velocity, the $\sigma_{v,\rm los}$ map is visually similar to the $V_{\rm rms}$ map. The covariance matrix for $V_{\rm rms}$ is obtained from combining the covariances of $v_{\rm los}$ and $\sigma_{v,\rm los}$ using the chain rule.

\subsection{Rotation and symmetry properties of the deflector \label{ssec:dynamical_properties}}

The stellar velocity measurements (bottom left panel of Fig.~\ref{fig:kinem_measur}) show signs of rotation, with a clear bipolar pattern whose orientation is consistent with the observed surface brightness. To quantify the degree of rotation, we compute the baryon projected specific angular momentum $\lambda_{\rm R}$ defined in \citet[see their Eq.~6]{2007MNRAS.379..401E}. Flux-averaging over the binned region we find $\lambda_{\rm R}=0.17$ (with 0.02 statistical uncertainty). This value of $\lambda_{\rm R}$ together with the observed ellipticity $\varepsilon$ serve as a useful classification between ``fast'' and ``slow'' (or nonregular) rotators \citet{2007MNRAS.379..401E,2016ARA&A..54..597C}. The classification of \citet{2016ARA&A..54..597C} state that galaxies with $\lambda_{\rm R} > \overline{\lambda_{\rm R}} \equiv 0.8 + \varepsilon/4$ can be classified as fast rotators (see also Knabel et al., in prep.). For \fifteenthirtyseven this would imply that the main deflector lies close the classification threshold with $\overline{\lambda_{\rm R}} = 0.13$, where we have used $\varepsilon = 1 - q_\ell=0.21$ and $q_\ell$ is the axis ratio measured in the \ir filter.

By comparing the rotational axis to the observed surface brightness, we can deduce the three-dimensional symmetry (prolate vs. oblate) of the deflector \citep[e.g.,][]{2006MNRAS.366..787K}. Using the dedicated Python software package \textsc{PaFit}\footnote{\url{https://pypi.org/project/pafit}}, we measure the kinematics position angle $\phi_{\rm kin}$ in the inner $0\farcs6$ region with mild binning (${\rm sS/N}=80$) to more finely sample the rotational axis. We find that $\phi_{\rm kin}$ is consistent with the surface brightness position angle $\phi_\ell$ measured in the \ir filter, with a difference of $\Delta\phi_{\rm kin} \equiv | \phi_{\rm kin} - \phi_\ell | = 13.6 \pm 4.5\ {\rm deg}$. Such an alignment indicates that the lensing galaxy has likely an oblate symmetry in three dimensions.

\section{Environment and line-of-sight contributions \label{sec:environment_and_los}}

\subsection{Perturbers \label{ssec:env_perturbers}}

We searched for galaxies that could cause lensing effects beyond convergence and shear terms. We start from our spectroscopic catalog combining MUSE and GMOS-S shown in Fig.~\ref{app:fig:environment} and detailed further in Sect.~\ref{app:sec:perturbers}. For each identified galaxies, we estimate the flexion shift $\Delta_{3}x$, which quantifies the shift in quasar image positions due to higher-order lensing effects \citep[see][]{McCully2017}. Following \citet{2019MNRAS.490..613S}, requiring $\log_{10}\Delta_{3}x > -4$ selects the four galaxies labeled G2, G3, G4 and G5 in Fig.~\ref{fig:color_image}, which we call perturbers.

We refine these perturbers mass estimates using our MUSE spectroscopic data. We directly measure the LoS velocity dispersion $\sigma_{v,{\rm los}}$ of G2, G3 and G4, as these have HST morphologies similar to the main deflector (G1) and their spectrum indeed is consistent with typical elliptical galaxy spectra. We therefore apply a similar approach as for G1 (Sect.~\ref{sec:kinem_measurements}), except that we measure $\sigma_{v,{\rm los}}$ within a circular aperture instead of a resolved map. We measure their effective (half-light) radius from fitting a Sérsic profile on the F160W filter from HST, which defines an aperture that probes a dynamically equivalent region for each galaxy \citep[e.g.,][]{2006MNRAS.366.1126C}. We integrate the MUSE spectrum within five times the effective radius, close to encompass their entire observed flux. We measure $\sigma_{v,{\rm los}}$ using a similar procedure as done for G1 in Sect.~\ref{sec:kinem_measurements}. Briefly, we perform the measurements in the same restframe wavelength range 3755-5400~\AA\ as for G1. We marginalize over additive polynomial of orders $\{8,10,12,14\}$ and multiplicative polynomial of orders $\{0,1\}$. We find on average that the measured velocity dispersions are $\approx1.3$ times smaller compared to our original catalog estimates (Sect.~\ref{app:sec:perturbers}). Assuming SIS profiles we obtain new estimates of their Einstein radius and flexion shift, on average $\approx1.7$ times smaller than the catalog estimates. After our refined measurements, only G2 and G3 have $\log_{10}\flexshift > -4$ at $>1\sigma$; however, we conservatively decide to include G4 in our lens models.

For G5, we must follow a different approach. From HST, the morphology and color of G5 differ from the other perturbers, and the MUSE data features strong emission lines but very weak continuum and absorption features. The data resolution and S/N being insufficient to measure a velocity profile. Therefore, we instead use the fitted relation of \citet{2019MNRAS.488.3143B} for star-forming galaxies (see their Table~J1) interpolated at $z=0.73$ to convert the stellar mass to a halo mass of $\log_{10}\left(M_{\rm peak}/M_{\odot}\right) \approx 11.7$. At this halo mass and redshift, a NFW profile with a concentration $c\approx5.7$ from \citet{2019ApJ...871..168D} requires baryons to produce an Einstein ring. Conservatively adding baryons as a point mass we obtain an Einstein radius of $\theta_{\rm E, G5}\approx0\farcs13$ ($\approx1.6$ smaller than the original catalog estimate). This smaller Einstein radius causes G5 to fail the flexion shift criterion at the $\sim1.3\sigma$ level, thus we do not include this galaxy in our fiducial lens models (see Sect.~\ref{sec:lens_modeling}), although we still run some model variations including it. We report the refined properties of all galaxy perturbers in Table~\ref{app:tab:perturbers_props}.

Our environment analysis also identifies a galaxy group candidate at the redshift of the main deflector; however, as detailed in Sect.~\ref{app:ssec:groups_identification}, this candidate group may be dynamically complex such that usual prescriptions to estimate its redshift, mass and centroid are expected to break down. While we cannot characterize the properties of the candidate group to warrant its inclusion via an additional mass component in our lens models, our analysis still includes first-order shear and convergence effects as well as higher-order terms from the most relevant group member candidates.

\subsection{External convergence \label{ssec:los_kappa_ext}}

We retrieve all sources within a $120\arcsec$ radius of the main deflector from the DELVE DR3 coadd catalog \footnote{\url{https://datalab.noirlab.edu/data/delve\#delve-dr3}} (Drlica-Wagner et al., in prep.), which provides $griz$ photometry derived from Dark Energy Camera imaging. Since DELVE DR3 does not supply photometric redshifts, we cross-match each source, within a tolerance of $0\farcs5$, to the photo-$z$ catalog of DECaLS DR10 \citep{DECALSDR10}, which is based on the same underlying imaging. For each matched source, we adopt the $i$-band-prior photo-$z$ estimate where available, falling back to the multi-band estimate otherwise. Sources without a valid photo-$z$ are discarded.

We apply the DELVE DR3 $5\sigma$ point-source detection limit in the $i$-band ($i_{\rm AUTO} \leq 23.5$) \footnote{Magnitudes in the AB system.}, and reject objects falling within regions manually masked around bright stars and obvious image artifacts. Star/galaxy separation is especially delicate here, as \fifteenthirtyseven\ lies at relatively low galactic latitude (${b=+20.3115^\circ}$) and the field is heavily populated by foreground stars that must be excluded from the LoS galaxy counts. It relies on the DELVE \texttt{ext\_mash} classifier, which combines morphological and photometric indicators into a discrete class from $0$ (high-confidence star) to $4$ (high-confidence galaxy). We further remove objects lying significantly below the $\mu_{\rm max}$--mag locus occupied by well-measured sources (fit iteratively with $3\sigma$ clipping over $16 < r < 20$), as these are dominated by stellar contaminants, blends, and low surface-brightness artifacts, a procedure similar to that described in \cite{Estrada23}.

To bracket the impact of residual stellar contamination on the inferred $\kext$, we carry forward three galaxy samples differing only in the stringency of the star/galaxy selection: (i) $\texttt{ext\_mash} \geq 3$ combined with the surface-brightness locus classification (128 galaxies), adopted as our fiducial sample; (ii) $\texttt{ext\_mash} = 4$ alone (146 galaxies), the most inclusive of genuine galaxies at the cost of some stellar contamination; and (iii) $\texttt{ext\_mash} = 4$ combined with the surface-brightness locus classification (115 galaxies), the most conservative. We repeat the $\kext$ inference for each of these three catalogs, and include the spread across the three as a systematic contribution to the final $\kext$ uncertainty.

To estimate $\kext$ from the catalogs, we make use of an updated version of the weighted number counts pipeline described in \cite{Wells_2023}. In summary, the number of galaxies within a $120\arcsec$ aperture around the lens galaxy is counted, and compared to many identical cutouts in random lines of sight within the DES catalog \citep{Abbott_2021}. We exclude all objects which lie within $5\arcsec$ of the lens, are explicitly modeled as perturbers\footnote{Our refined mass estimate for G5 (the least massive perturber, see Sect.~\ref{ssec:env_perturbers}) was performed after constructing LoS catalog, thus it was not excluded for inferring \kext. However, removing a single low-mass object from the input catalog is not expected to affect our results beyond our total uncertainties.} or fall below a 23-magnitude cutoff imposed to ensure completeness. We also store weights by the photometric redshift, inverse distances and potentials of these galaxies. Following \cite{Wells_2023} and its application in other TDCOSMO analyses \citep[e.g.][]{2026arXiv260414145S}, simulated lines of sight are then sampled according to the resulting distribution of relative weighted number counts. 

In our updated analysis, these simulated lines of sight are taken from the Euclid Flagship simulation, a state of the art 4-trillion particle N-body simulation, populated with 3.4 billion galaxies, and with weak lensing convergence values stored in 800 million pixels in 200 redshift shells \citep{Castander_2025}, accessed with CosmoHub \citep{TALLADA2020100391,2017ehep.confE.488C}. To estimate $\kappa_\mathrm{ext}$ from the sampled lines of sight, we first determine $\kappa_\mathrm{s}$, $\kappa_\mathrm{d}$ via linear interpolations of adjacent redshift shells, and $\kappa_{\mathrm{ds}}$ via a discrete integral using Eq.~3.1 in \cite{Johnson_2025}. These quantities are then combined according to Eq.~\ref{eq:los_breakdown}. A complete description of this method will be presented in Johnson et al. (in prep.).

\begin{figure}
    \centering
    \includegraphics[width=\linewidth]{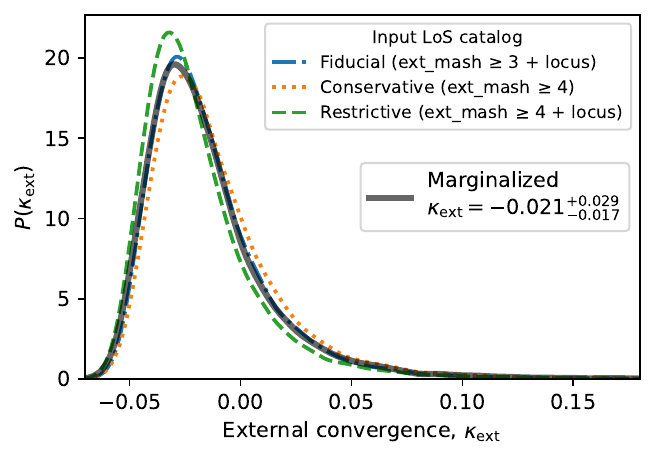}
    \caption{Posterior distribution of external convergence $\kext$ for \fifteenthirtyseven, that is the effective contribution of LoS objects other than the main deflector and perturbers to the net lensing effect (see Sect.~\ref{ssec:los_kappa_ext} for details on the input catalogs).}
    \label{fig:kappa_ext_histo}
\end{figure}

The resulting $\kappa_\mathrm{ext}$ posterior distributions are shown in Fig.~\ref{fig:kappa_ext_histo} for each of the three input catalogs. For each catalog, the number counts and external convergence indicate a somewhat under-dense lens environment with a slightly negative median. We marginalize over the three histograms assuming equal weights to obtain our final posterior, giving $\kext = -0.021^{+0.029}_{-0.017}$.

\section{Strong lensing constraints \label{sec:lens_modeling}}

\subsection{Lens modeling assumptions \label{ssec:lens_model_general}}

We use the lens modeling software packages \glee \citep{SuyuHalkola2010, Suyu+2012} and \lenstro \footnote{\url{https://github.com/lenstronomy/lenstronomy} (v. 1.11.7).} \citep{Birrer2018lenstro,2021JOSS....6.3283B} to model the HST data of \fifteenthirtyseven. Modeling the same data with different software is an efficient way to probe and quantify potential systematic biases in strong lensing analyses, as shown by \citet{2024A&A...692A..87G}. We performed in \citet{2022A&A...667A.123S} such a comparative analysis for the quadruply lensed quasar ${\rm WGD}\,2038-4008$, which further improved our understanding of modeling systematics and informed subsequent analyses. Due to specific lens modeling requirements in the context of time-delay cosmography, conducting such comparative analyses significantly increases the complexity and duration of the lens modeling step, hence it cannot be realistically conducted for all systems in our growing sample. Nevertheless, it remains important to reiterate the exercise with different lensing systems, updated versions of the codes and even different software packages.

The two lens modeling teams (led by A.G. for \lenstro and A.G.S. for \glee) remained blind to each others constraints on key parameters during the lens modeling stage. The two teams agreed on describing the total mass of the main deflector by a single power-law profile embedded in a uniform external shear field\footnote{There is no guarantee that such a uniform shear field only captures the shear that is external to the main deflector \citep{2024MNRAS.531.3684E,2026arXiv260424908B}. We nevertheless retain the ``ext'' in parameter names.}. The two teams also agreed on the list of explicitly modeled perturbers (redshifts, positions and mass profile). Both teams use the measured time delays as extra constraints during lens modeling, adding the (modeled) time-delay distance $\Ddt^{\rm model}$ as an extra model parameter \footnote{Alternatively, an equivalent $\Hc^{\rm model}$ parameter can be directly sampled (see Eq.~\ref{eq:weights_prior_ddt_hc} for a necessary re-weighting step in this case).}. In contrast, the two teams did not necessarily follow the same assumptions for other aspects of their models. For instance, the surface brightness of the main deflector, the quasar images and the quasar hosts were modeled differently. Other differences were the choice of pixels to exclude from the imaging likelihood, and the choice of pixels with boosted uncertainties. We give in-depth descriptions on the modeling process in Sect.~\ref{app:sec:lens_model_details} for each code. In particular, we list and discuss in Sect.~\ref{app:ssec:model_comp} the main similarities and differences between the \lenstro and \glee lens models.

As noted in previous studies \citep[e.g.,][]{Schneider_2013,Sonnenfeld2018,Kochanek2020}, certain choices of mass profiles are known to artificially break the MSD, by restricting the radial behavior of the mass distribution. Only for the lens modeling step, we choose to break the MSD through the assumption of a single power-law profile during lens modeling; subsequently, we allow for an MSD transformation with parameters $\lint$ and $\kext$ on the mass density profile of the main deflector, and we will use stellar dynamics and LoS information to further constrain $\lint$ and $\kext$ (see Sect.~\ref{sec:dynamical_weighting}). At the lens-modeling stage, the model parameter $\Ddt^{\rm model}$ therefore represents the time-delay distance uncorrected for the MSD.

Regarding the treatment of perturbers G2, G3 and G4, the small redshift differences compared to G1 and our galaxy group identification step both indicate that these galaxies may belong to the same group at $z\approx0.59$. As such, we place all these galaxies on the same redshift plane at $z=\zd=0.590$ in our lens models. We also include in our model variations a set of multi-plane models with perturber G5 at its own redshift ($z=0.730$), as additional systematic tests. As mentioned in Sect.~\ref{ssec:env_perturbers}, we did not select any galaxy group to be included explicitly our lens models. We expect G2, G3 and G4 (which are potential members of the $z\approx0.59$ group candidate) combined with the external shear component to provide sufficient flexibility to capture higher-order lensing effects caused by the presence of group-scale halo at the position of the multiple images.

\subsection{Bayesian model averaging \label{ssec:bic_weighting}}

Within each modeling code, we compare and combine a set of model variations in a Bayesian framework. We approximate the Bayesian evidence of each model using the BIC, similarly to our kinematics measurements (Sect.~\ref{sec:kinem_measurements}). For each model, we compute the BIC and corresponding weight $w_{\rm BIC} = \exp\!\left[-\Delta {\rm BIC}/\,2\right]$, where $\Delta {\rm BIC} = \left({\rm BIC} - \rm BIC_{min}\right)$. However, this weight does not account for the finite and limited number of model variations we can possibly run, as well as the inherent uncertainty on the BIC value itself, and thus the uncertainty on $\Delta {\rm BIC}$. Therefore, we follow past cosmographic analyses \citep[e.g.][]{Birrer2019,2020MNRAS.493.4783Y,2022A&A...667A.123S} to estimate an error term $\sigma_{\Delta \rm BIC}$ from a set of model variations (see also Sects~\ref{app:ssec:lenstronomy_modeling} and \ref{app:ssec:glee_modeling} for the descriptions \glee and \lenstro models, respectively). We then convolve the original weights distribution with a gaussian kernel $h(\,\cdot\,,\,\sigma_{\Delta \rm BIC})$ to obtain the corrected distribution:
\begin{align}
\label{eq:bic_weights}
    w_{\rm BIC}^\ast &= h({\rm BIC}_n,\,\sigma_{\Delta \rm BIC}) \ast w_{\rm BIC} \ .
\end{align}
We employ the weights $w_{\rm BIC}^\ast$ to obtain joint posterior distributions that marginalize over all model variations for a given modeling software.

\subsection{Results from lens modeling \label{ssec:lens_model_posterior}}

\begin{figure*}[!ht]
    \centering
    \includegraphics[width=0.9\linewidth]{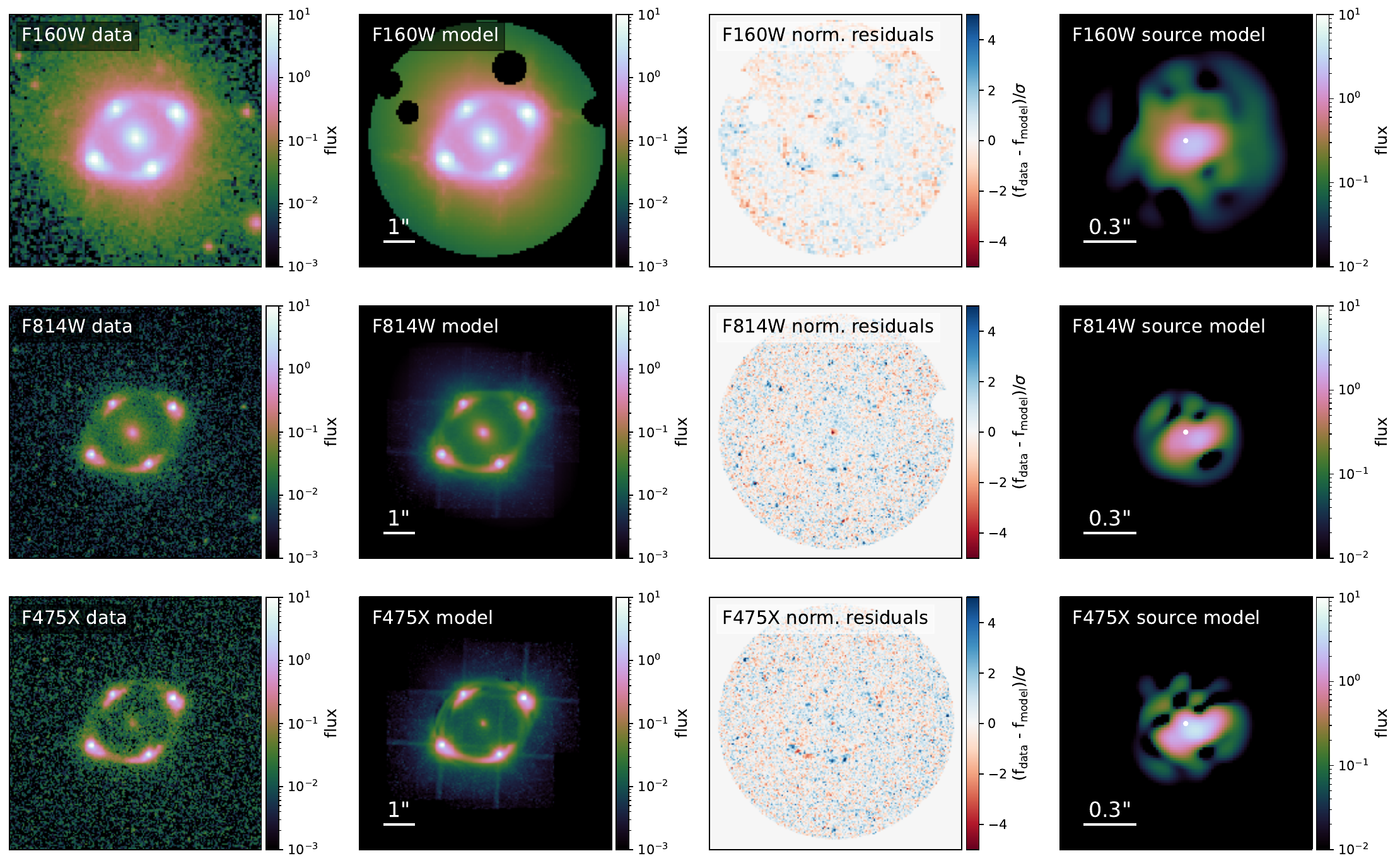}
    \\
    \vspace{-3.5mm}
    \rule{0.95\linewidth}{1pt}
    \\
    \includegraphics[width=0.9\linewidth]{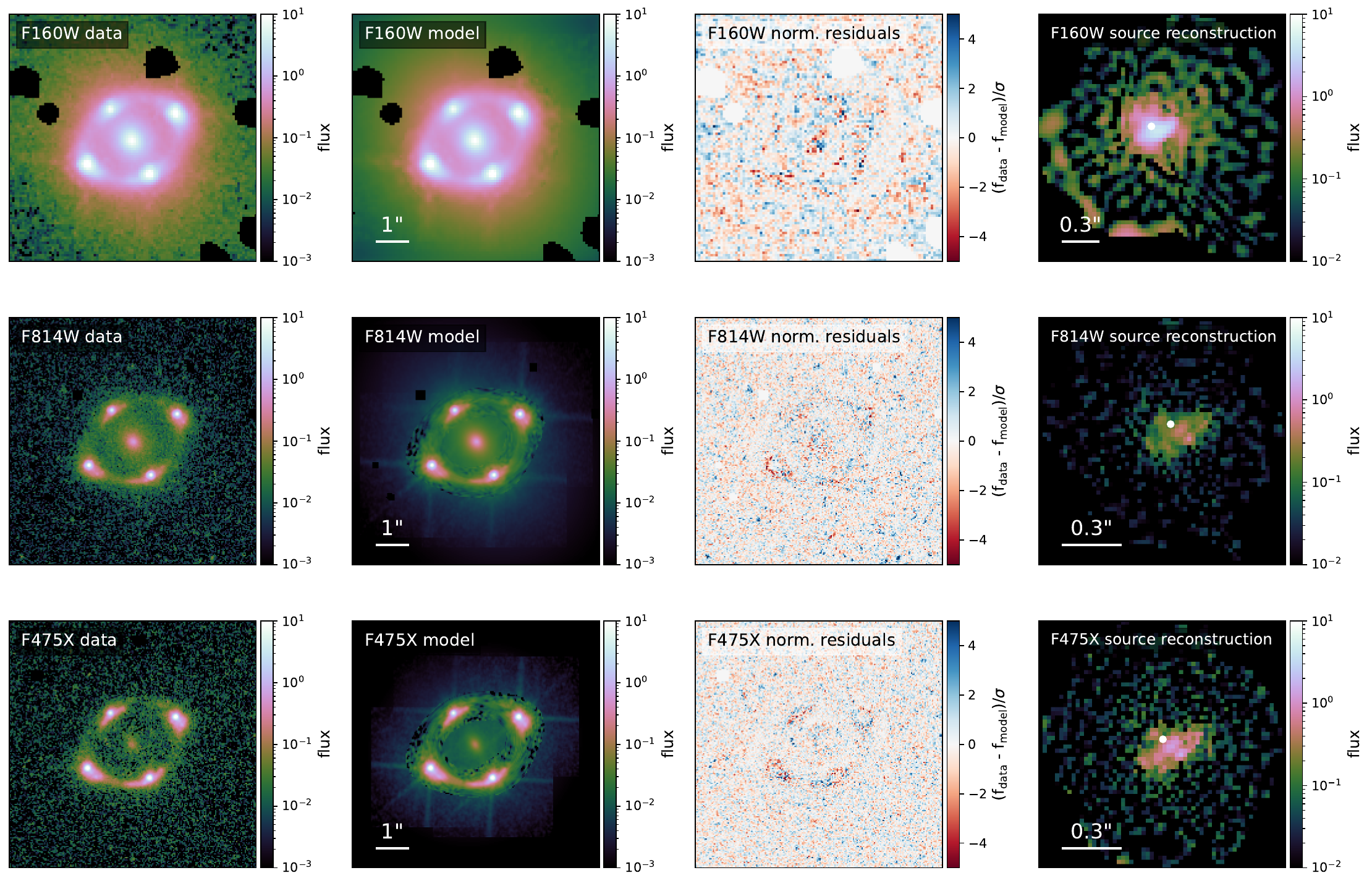}
    \caption{Best-fit lens model obtained with \lenstro (top three rows) and \glee (bottom three rows). Each row shows one HST filter, and each column shows the imaging data, the corresponding model, normalized residuals and the quasar host model (the white dot indicates the position of the quasar). We see no statistically significant residual patterns, showing that all lens models reproduce well the observations given the modeled uncertainties. Note that the image and source plane field of view slightly differ between \lenstro and \glee models.}
    \label{fig:bestfit_lens_models}
\end{figure*}

\begin{figure*}[!ht]
\sidecaption
    \includegraphics[width=0.7\linewidth]{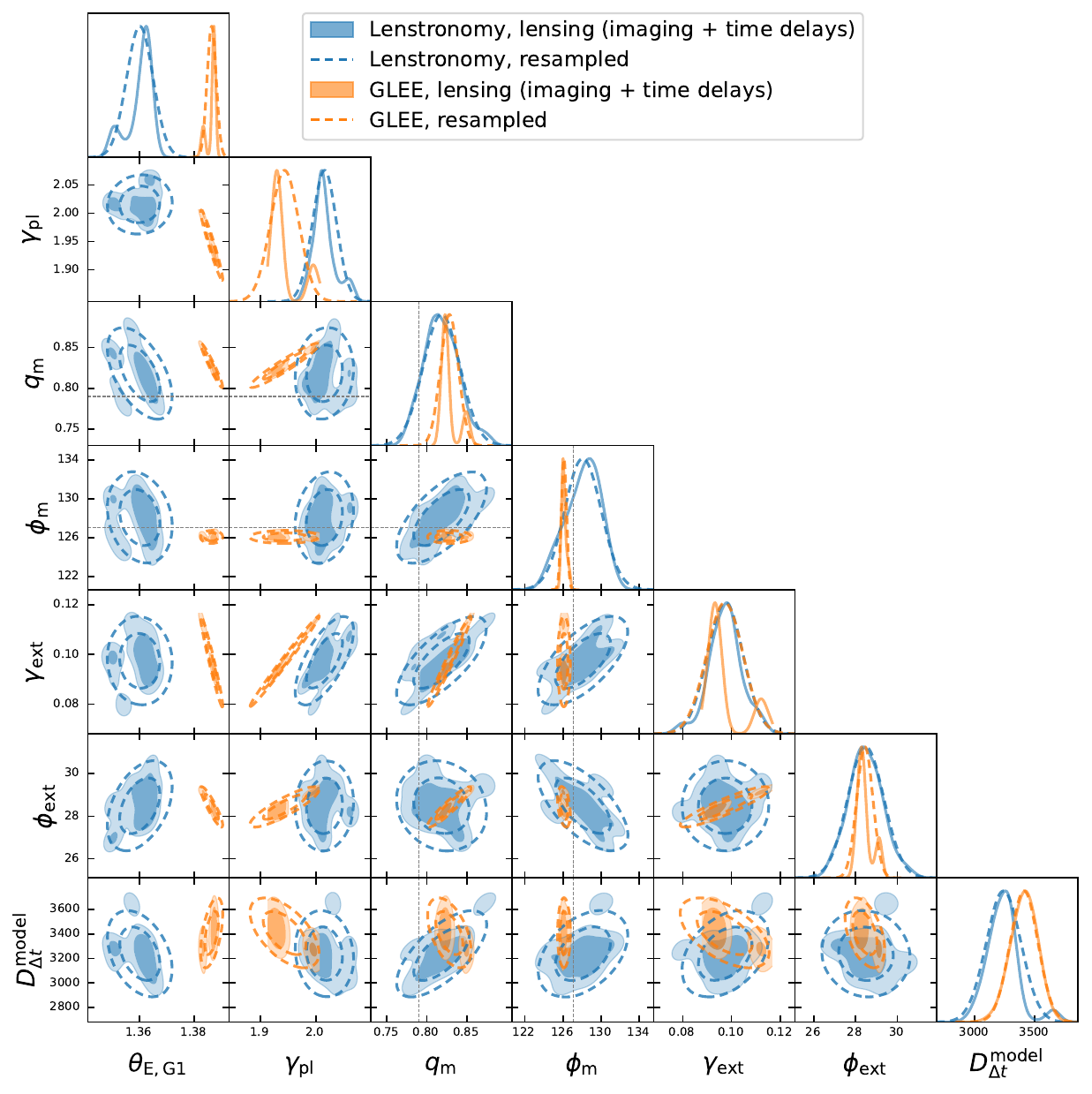}
    \caption{Joint posterior distribution of a subset of key parameters from \lenstro and \glee lens models, after marginalization over various model variations for systematic errors. From left to right, the parameters are the main deflector Einstein radius $\theta_{\rm E, G1}$, the mass density slope $\gamma_{\rm pl}$ the total mass axis ratio and position angle ($q_{\rm m}$ and $\phi_{\rm m}$), the external shear parameters ($\gamma_{\rm ext}$ and $\phi_{\rm ext}$) and the MST-uncorrected time-delay distance $\Ddt^{\rm model}$. For each model separately, we improve parameter space sampling by fitting a multivariate Gaussian model and resample the distributions before including kinematics and LoS constraints in Sect.~\ref{sec:dynamical_weighting}. The thin dashed lines plotted over $q_{\rm m}$ and $\phi_{\rm m}$ panels show the corresponding best-fit values from the surface brightness model (see Sect.~\ref{app:ssec:lens_light_model} for details).}
    \label{fig:lens_model_posterior}
\end{figure*}

We show in Fig.~\ref{fig:bestfit_lens_models} the fit to the HST imaging obtained with \lenstro and \glee, respectively, where we have chosen the best-BIC model for each code. We refrain from quoting corresponding $\chi^2$ values as they cannot be compared between the two codes due to their different treatment of pixel uncertainties (see Sect.~\ref{app:ssec:model_comp}). However, within a given code, individual model variations can be compared through their BIC values, from which we can weight their individual contributions (Tables~\ref{app:tab:lenstro_model_variations} and \ref{app:tab:glee_model_variations}).

We show in Fig.~\ref{fig:lens_model_posterior} the joint posterior distributions of a subset of lens model parameters. Overall \lenstro and \glee models broadly agree for most of the key parameters. External shear parameters are among the most consistent parameters. The time-delay distance $\Ddt^{\rm model}$, which encapsulates the \Hc dependence, agree to within $1\sigma$. While most parameters are consistent within $1-2\sigma$ level, we note differences in posterior widths and a $\sim 3\sigma$ shift G1 Einstein radius. Independent teams conducted the lens modeling step of this analysis and the input data for each code was not fully identical in terms of field of view, pixel masking and handling of pixel uncertainties. These differences are expected to affect the widths of the posteriors, primarily because the the treatment of pixel-level uncertainties (i.e., noise map modeling), differ between \lenstro and \glee models. These modeling differences are expected to affect the width of the posterior distribution, in turns affecting their apparent statistical consistency. We discussed further in Sect.~\ref{app:ssec:model_comp} the similarities and differences between the two codes. Nevertheless, the uncertainty on $\Ddt^{\rm model}$ remains similar, as it is also driven by the measured time delays.

The discrepancy between G1 Einstein radius values is cause by degeneracies between the main deflector and nearby perturbers. Most of \glee lens model variations favor perturbers (G2, G3, G4) that have significantly lower mass than their measured priors from MUSE. To maintain the total Einstein radius---which the most robust quantity constrained by strong lensing---, the Einstein radius of G1 is pushed upward. On the contrary, in the \lenstro model, the perturbers masses remain consistent with their priors and a lower value of $\theta_{\rm E, G1}$ is favored. These lens model differences are marginalized over after including the effect of the MST constrained with stellar kinematics and LoS information, as shown in Sect.~\ref{sec:dynamical_weighting}.

\section{Strong lensing, stellar kinematics and line of sight to infer cosmology \label{sec:dynamical_weighting}}

\subsection{Mass-sheet parametrization \label{ssec:msd_parametrization}}

The internal part of the MSD alters the density profile of the main deflector, in particular along the radial direction. We relax the assumption of a single power-law density profile by introducing an internal mass-sheet parameter $\lint$ that transforms the density profile via the MST. However, we must ensure the physicality of the transformed profile. We do so by adding a mass component designed to mimic a pure mass sheet over the inner region constrained by lensing while returning to zero at larger radii. We achieve this behavior with the following cored density profile \citep[][]{Blum2020,Birrer2020_TDC4,2023A&A...673A...9S}:
\begin{align}
    \label{eq:def_kappa_cored_profile}
    \kappa_{\rm c}(\imagepos; R_{\rm c}) = \frac{R_{\rm c}^2}{R_{\rm c}^2  + \boldsymbol{\theta}^2} \ ,
\end{align}
where $R_{\rm c}$ is the core radius. For $R_{\rm c}$ sufficiently large, Eq.~\ref{eq:def_kappa_cored_profile} indeed behaves like a uniform sheet of mass with amplitude unity. By scaling $\kappa_{\rm c}$ by $(1 - \lint)$, we can identify $\lint$ as an approximate mass-sheet parameter following the definition of the MST (Eq.~\ref{eq:mst}). We can then transform the convergence $\kappa_{\rm model}$ constrained from lens modeling as follows:
\begin{align}
    \label{eq:kappa_int_mst}
    \kappa_{\rm model}'(\imagepos) = \lint\kappa_{\rm model}(\imagepos) + (1 - \lint)\kappa_{\rm c}(\imagepos; R_{\rm c})\ .
\end{align}
$\kappa_{\rm model}'$ is more flexible than $\kappa_{\rm model}$ in a way that is maximally degenerate with \Hc by construction, and that can be constrained with stellar kinematics measurements. We set a prior on \lint by requiring that (1) $\kappa_{\rm model}'$ should remain positive and decrease monotonically (upper bound on \lint) and (2) the three-dimensional mass of the cored component $\kappa_{\rm c}$ should not exceed the total mass of a Navarro Frenk \& White \citep[NFW;][]{Navarro1997} profile within the same volume \citep[see][]{Birrer_2020}. For \fifteenthirtyseven we find that a uniform prior $\lint \sim \mathcal{U}(0.8, 1.2)$ fulfills these requirements. To ensure the desired mass-sheet behavior of \lint, we place a prior on $R_{\rm c}$ with bounds set to $5\times\thetaE$ and $10\times\thetaE$ respectively \citep[][]{Birrer2020_TDC4,2026arXiv260414145S}, giving $R_{\rm c} \sim \mathcal{U}(6\farcs6, 13\farcs1)$ \footnote{We used $\thetaE=1\farcs31$ based on preliminary lens models; the $R_{\rm c}$ prior range only marginally changes given our final \thetaE estimates.}.

Multiplying $\kappa_{\rm model}'$ by $(1 - \kext)$ obtained in Sect.~\ref{sec:environment_and_los} effectively corrects for external contributions to the modeled convergence \citep{2023A&A...673A...9S}. The density profile of only the main deflector is then 
\begin{align}
\nonumber
\label{eq:kappa_full_mst}
    \kappa_{\rm gal}(\imagepos) &= (1 - \kext)\kappa_{\rm model}(\imagepos)' \\
    &= (1 - \kext) \big[ \lint\kappa_{\rm model}(\imagepos) + (1 - \lint)\kappa_{\rm c}(\imagepos; R_{\rm c}) \big] \ ,
\end{align}
where we identify the total MST parameter $\lambda\equiv(1-\kext)\lint$ already defined in Eq.~\ref{eq:mst_decomp}. Since the MSD also affects the time-delay distance (Sect.~\ref{ssec:mst_theory}), we must transform our lensing-constrained parameter $\Ddt^{\rm model}$ to obtain the actual time-delay distance 
\begin{align}
\label{eq:ddt_full_mst}
    \Ddt = \frac{\Ddt^{\rm model}}{\lambda} \ .
\end{align}

At this point our updated lens model explicitly incorporates the internal \lint and external \kext contributions to the MSD constrained by stellar kinematics and by the LoS analysis, respectively. Next we present the rest of the dynamical model we use to predict stellar kinematics along with necessary additional model parameters and priors.

\subsection{Dynamical model \label{ssec:dynamical_params}}

We predict the stellar kinematics of the main deflector using Jeans anisotropic modeling \citep[JAM;][]{2008MNRAS.390...71C,2020MNRAS.494.4819C}. This model requires different ingredients: axisymmetric models of the projected mass density and surface brightness of the galaxy, inclination angle, stellar anisotropy and black hole mass. For the mass density, we use our posterior distributions on $\kappa_{\rm model}$ from lens modeling, which we transform to a posterior distribution on $\kappa_{\rm gal}$ using Eq.~\ref{eq:kappa_full_mst} and $R_{\rm c}$, \lint and \kext parameters sampled from their respective priors. From $\kappa_{\rm gal}$ we obtain the surface mass density in physical units using
\begin{align}
    \label{eq:def_sigma_gal}
    \Sigma_{\rm gal} = \kappa_{\rm gal}\,\Sigma_{\rm crit} \ ,
\end{align}
where the critical surface density $\Sigma_{\rm crit}$ is a function of cosmological distances:
\begin{align}
\label{eq:sigma_crit_ddt_dd}
    \Sigma_{\rm crit} = \frac{c^2}{4\pi G}\frac{\Ds}{\Dd\Dds} = \frac{c^2}{4\pi G} \frac{\Ddt}{(1 + \zd)\,\Dd^2} \ ,
\end{align}
where we have used the definition of \Ddt (Eq.~\ref{eq:def_ddt}), and $c$ is the speed of light and $G$ the gravitational constant. Equation \ref{eq:sigma_crit_ddt_dd} effectively makes the link between constraints from lensing imposed on $\Ddt^{\rm model}$ and transformed with Eq.~\ref{eq:ddt_full_mst}, and those from stellar kinematics and external convergence.

For the surface brightness $\Sigma_{\ell}$, we aim at maximizing the overlap with the wavelength range used for kinematics measurements. We proceed similarly as \citet{2026A&A...707A.314S} and construct a stitched surface brightness model $\Sigma_\ell$ constrained by both \uvisone and \ir filters, in the inner and outer region respectively (see Sect.~\ref{app:ssec:lens_light_model} for details). We note that $\Sigma_\ell$ is very well constrained by the HST data, such that its uncertainty is subdominant compared to the other components of the dynamical model. Therefore, we fix the parameters defining $\Sigma_\ell$ to their best-fit values to improve sampling efficiency.

We restrict the inclination angle $i$ to values that are consistent with the observed (projected) axis ratio $q_\ell$ of the light distribution. We proceed as \citet{2026arXiv260414145S} and the results from \citet{2018ApJ...863L..19L} based on a sample of early-type galaxies from the SDSS-IV DR14 MaNGA survey to define a normal prior on the intrinsic three-dimensional axis ratio $q_{\ell, \rm intr} \sim \mathcal{N}(0.74, 0.08)$.

Stellar orbits in JAM can be described using spherically-aligned or cylindrically-aligned velocity ellipsoids. Given that the deflector lies in between slow and fast rotator categories (Sect.~\ref{ssec:dynamical_properties}), we follow the recommendations of \citet{2020MNRAS.494.4819C} and compare both extreme solutions to the Jeans equations. With the cylindrically-aligned solution, the stellar anisotropy parameter is defined as $\beta_{\rm ani} = 1 - \left( \sigma_z / \sigma_R \right)^2$ \citep{2008MNRAS.390...71C}, where $\sigma_z$ and $\sigma_R$ are the stellar velocity dispersions along the rotational and radial axes of the galaxy in cylindrical coordinates, respectively. We adopt a uniform prior on $\left( \sigma_z / \sigma_R \right) \sim \mathcal{U}(0, 1)$, based on observations of rotating elliptical galaxies \citep[e.g.,][]{2016ARA&A..54..597C}. In the spherically-aligned solution, the anisotropy parameter is instead $\beta_{\rm ani} = 1 - \left( \sigma_\theta / \sigma_r \right)^2$, where $\sigma_\theta$ and $\sigma_r$ are the tangential and radial stellar velocity dispersions in spherical coordinates, respectively. We adopt a Gaussian prior on $\left(\sigma_\theta / \sigma_r \right) \sim \mathcal{N}(1, 0.07)$, derived a sample of 13
representative early-type galaxies from \citet{2026enap....4..122C}. For both orbit alignment cases, we assume a radially constant $\beta_{\rm ani}$. Recent works have shown that a constant anisotropy is a better description of observations \citep{2026enap....4..122C} and simulations \citep{2026ApJ..1000..264V}, as opposed to the commonly used Osipkov-Merritt model \citep{1979SvAL....5...42O,1985AJ.....90.1027M}.

The resolution of our stellar kinematics maps is not sufficient to constrain the mass of the central black hole of the lens. However, we cannot simply ignore its influence on the predicted kinematics maps as shown by \citet{2025A&A...701A.280W}. Therefore, similar to \citet{2026arXiv260414145S}, we set an informative prior on the mass of the black hole. In particular, we use the $M_{\rm BH}-\sigma$ relation from \citet[][Table 3]{2013ApJ...764..151G} to convert the stellar velocity dispersion to a black hole mass $M_{\rm BH}$ for \fifteenthirtyseven. For the velocity dispersion, we consider the inverse-variance weighted average over the bins we extracted in Sect.~\ref{sec:kinem_measurements} ($\sigma=243 \pm 3\ {\rm km\,s^{-1}}$). Applying the formula of \citet{2013ApJ...764..151G} specific to their sample of 28 elliptical galaxies, we find $\log_{10}\left(M_{\rm BH}/M_{\rm \odot}\right) \sim \mathcal{N}(8.6, 0.3)$, which we use as a prior in our dynamical model\footnote{After adding the black hole component one should subtract off the corresponding lensing mass to conserve the Einstein radius. However, the black hole mass here is prior-dominated and remains negligible compared to that of the main deflector ($\sim5\times10^{11}\ {\rm M}_\odot$) such that this step is unnecessary in our case (effect $<0.5\%$).}. To assess the sensitivity of our results to this prior, we also used a prior twice as large and found no noticeable change in key posterior distributions, including \Hc and the MST parameters.

We use the software package \textsc{jampy} \footnote{\url{https://pypi.org/project/jampy/}} \citep{2008MNRAS.390...71C} to predict the $V_{\rm rms}$ map from the joint distribution of parameters. \textsc{jampy} expects the mass density and surface brightness of the galaxy to be expressed in terms of concentric elliptical Gaussian profiles, with joint position angles. Therefore, we perform a multi-Gaussian expansion (MGE) of both the convergence ($\kappa_{\rm gal}$ from Eq.~\ref{eq:kappa_full_mst}) and the surface brightness $\Sigma_\ell$. We use \textsc{mgefit}\footnote{\url{https://pypi.org/project/mgefit/}} \citep{2002MNRAS.333..400C} to decompose the corresponding (one-dimensional) spherical profiles. For each MGE we use 20 Gaussians, which ensures that the relative error on the decomposed profile is less than a percent over the relevant radial range. We fold back the ellipticity information separately for the mass and surface brightness by setting the axis ratio of all Gaussians to that of the original model. The position angle of the mass density is required to match that of the surface brightness, which are statistically consistent from our lens modeling results (see dashed lines in Fig.~\ref{fig:lens_model_posterior}).

We fold the stellar kinematics and LoS constraints by importance sampling the lensing posterior samples obtained in Sect.~\ref{sec:lens_modeling} and shown in Fig.~\ref{fig:lens_model_posterior}. More specifically, for each sample of the combined chain we compute the weights associated to the stellar kinematics likelihood, assumed a multi-variate Gaussian distribution defined by the measured $V_{\rm rms}$ and covariance matrix obtained in Sect.~\ref{sec:kinem_measurements} and given the prior distributions described above. We similarly impose our prior on the intrinsic axis ratio by multiplying the weights by its corresponding Gaussian distribution.

\subsection{Cosmological model \label{ssec:cosmo_model}}

We evaluate Eqs.~\ref{eq:ddt_full_mst} and \ref{eq:sigma_crit_ddt_dd} by sampling \Ddt and \Dd from the source and deflector redshifts assuming a cosmological model. We assume a flat \lcdm cosmology parametrized by \Hc and \Om. We use a prior $\Om\sim\mathcal{N}(0.334, 0.018)$ obtained from the Pantheon+ Type Ia supernovae data \citep{2022ApJ...938..110B}. For \Hc, the primary goal of this analysis, we use a broad uniform prior $\Hc \sim \mathcal{U}(30, 110)\ {\rm km\,s^{-1}\,Mpc^{-1}}$. The mild $\Hc-\Om$ covariance (Fig.~\ref{app:fig:dynamical_model_full_posterior}) makes our result robust to any choice of \Om prior that is similarly precise as Pantheon+ \citep[e.g., see also][]{2025A&A...704A..63T,2026arXiv260414145S}.

\subsection{Results from the combined constraints \label{ssec:joint_results}}

\begin{figure*}[!ht]
\centering
    \mbox{
    \adjincludegraphics[width=0.7\linewidth,valign=t]{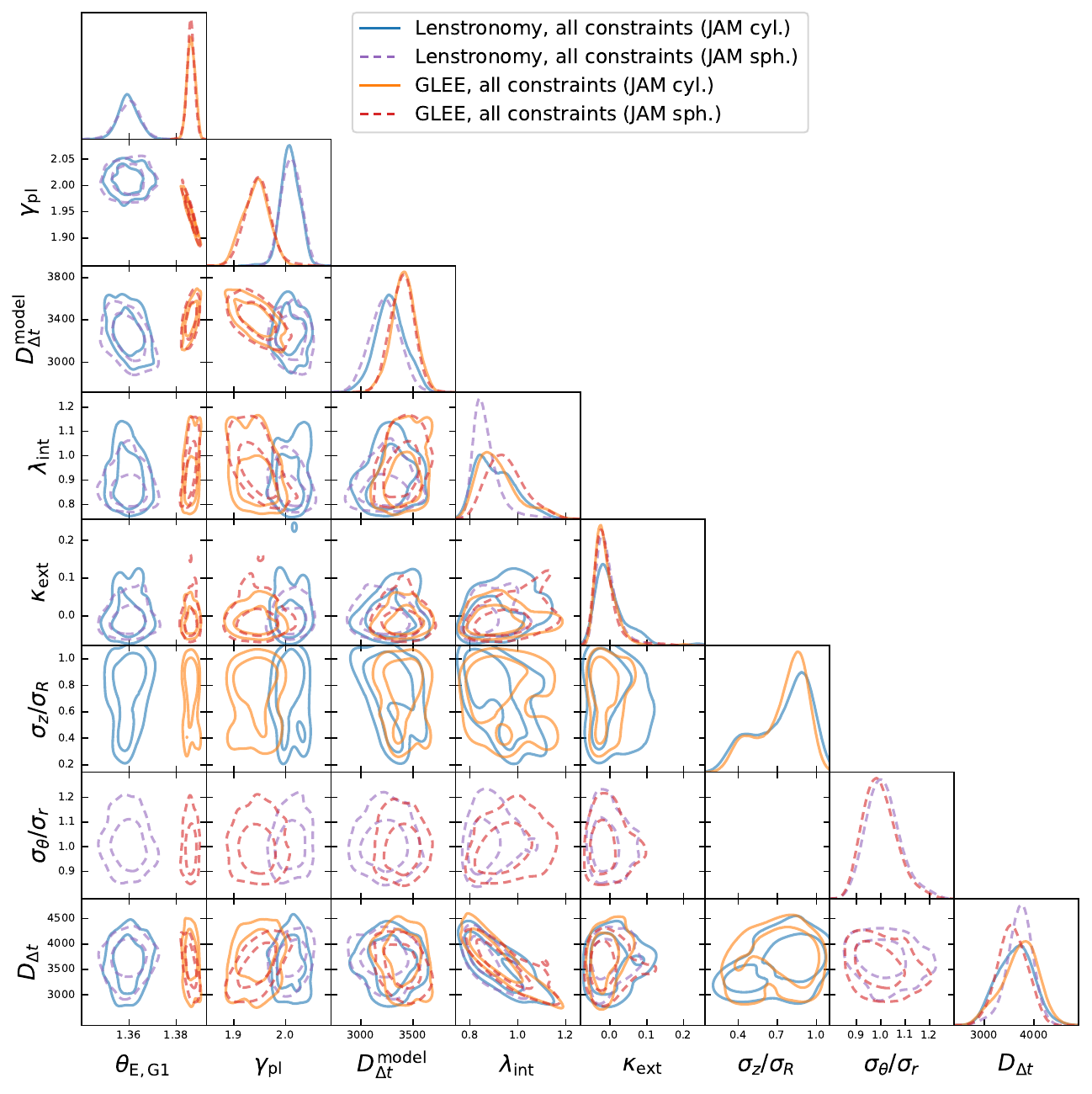}
    \adjincludegraphics[width=0.24\linewidth,valign=t]{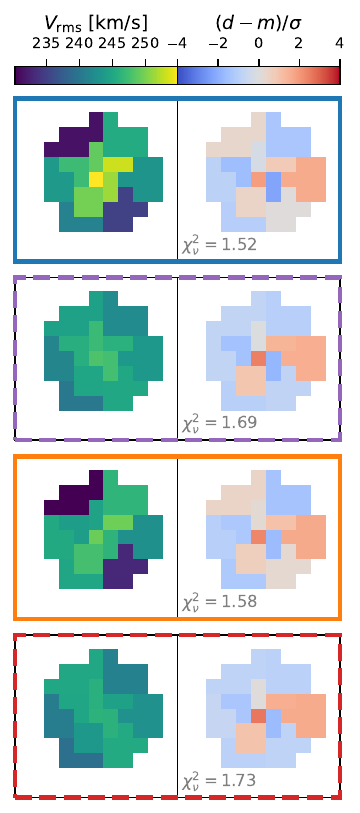}
    }
    \caption{\textit{Left panel}: joint posterior distribution of key parameters of our combined lensing, dynamics, LoS and cosmological inference (see Sect.~\ref{sec:dynamical_weighting} for details and Fig.~\ref{app:fig:dynamical_model_full_posterior} for the full parameter space). Distributions are shown for each lens model (\lenstro and \glee) and each choice of JAM parametrization (cylindrically or spherically aligned stellar orbits) \textit{Right panel}: best-weighted dynamical models and reduced residuals corresponding to the same four models (the observed $V_{\rm rms}$ map is shown in Fig.~\ref{fig:kinem_measur}).
    }
    \label{fig:dynamical_model_posterior}
\end{figure*}

Our final constraints including strong lensing, stellar kinematics, LoS and cosmology are shown in the left panel of Fig.~\ref{fig:dynamical_model_posterior}. The right panel of that figure shows the corresponding fit to the stellar kinematics measurements. Both JAM parametrizations lead to similar constraints on model parameters, although the cylindrically aligned ellipsoids result in lower reduced $\chi^2$ values for both \glee and \lenstro lens models. The radial flexibility in the mass profile introduced with internal and external MST parameters brings the two lens models  back into agreement by enlarging uncertainties and preferring slightly different \lint and orbital anisotropy distributions. The \lint distributions are consistent with unity at the $1\sigma$ level (mild MST corrections to the power-law mass profile), although the distributions are skewed towards $<1$ vales (i.e., slightly negative internal mass sheet). The anisotropy in both JAM parametrizations is consistent with zero ($\beta_{\rm ani}\approx0$), although the less informative prior on $\sigma_z/\sigma_{\rm R}$ compared to on $\sigma_\theta/\sigma_r$ allows for a slightly larger anisotropy. We note that the slightly steeper slope from the \lenstro lens models enables that model to reach a larger velocity dispersion, more consistent with the innermost spaxel measurement. In Table~\ref{tab:constraints_summary} we summarize constraints on key quantities inferred throughout our analysis.

\subsection{Fiducial model for population-level inferences \label{ssec:choice_for_hierarc}}

The lensing-only constraints (imaging + time delays) presented in this work will be included in our time-delay lens sample used in population-level inferences of \Hc (TDCOSMO Collaboration, in prep.). More specifically, our hierarchical inference scheme uses the joint posterior distributions of the main deflector's Einstein radius, mass density and axis ratio as well as the lensing-informed time-delay distance ($\Ddt^{\rm model}$). Our lensing-only constraints for \fifteenthirtyseven have been obtained with both \lenstro and \glee. However we cannot simply marginalize over them because the posterior distributions over the parameters of interest are not all statistically consistent, in particular $\theta_{\rm E,G1}$, due to the reasons detailed in Sect.~\ref{ssec:lens_model_posterior}.

We consider three arguments to motivate the use the \lenstro lens model of \fifteenthirtyseven for subsequent hierarchical inferences. First, the G1 Einstein radius is expected to be more accurate than from \glee, as it is degenerate with the perturbers Einstein radii and these remain consistent with their measured velocity dispersions only in the \lenstro model. Second, the position angle of G1's mass density is more consistent with its surface brightness position angle (see dashed lines in Fig.~\ref{fig:lens_model_posterior}), ensuring self-consistency during axisymmetric dynamical modeling. Third, the \lenstro lens model leads to a slightly better fit to the stellar kinematics measurements in terms of $\chi^2$ (see Fig.~\ref{fig:dynamical_model_posterior}). Therefore, we will consider the \lenstro lensing constraints in our next hierarchical inferences. We will still consider the \glee for assessing any systematic shift in inferred parameters at the sample level.

\section{Hubble constant measurement \label{sec:cosmo_measurements}}

\begin{figure}[!ht]
    \centering
    \includegraphics[width=\linewidth]{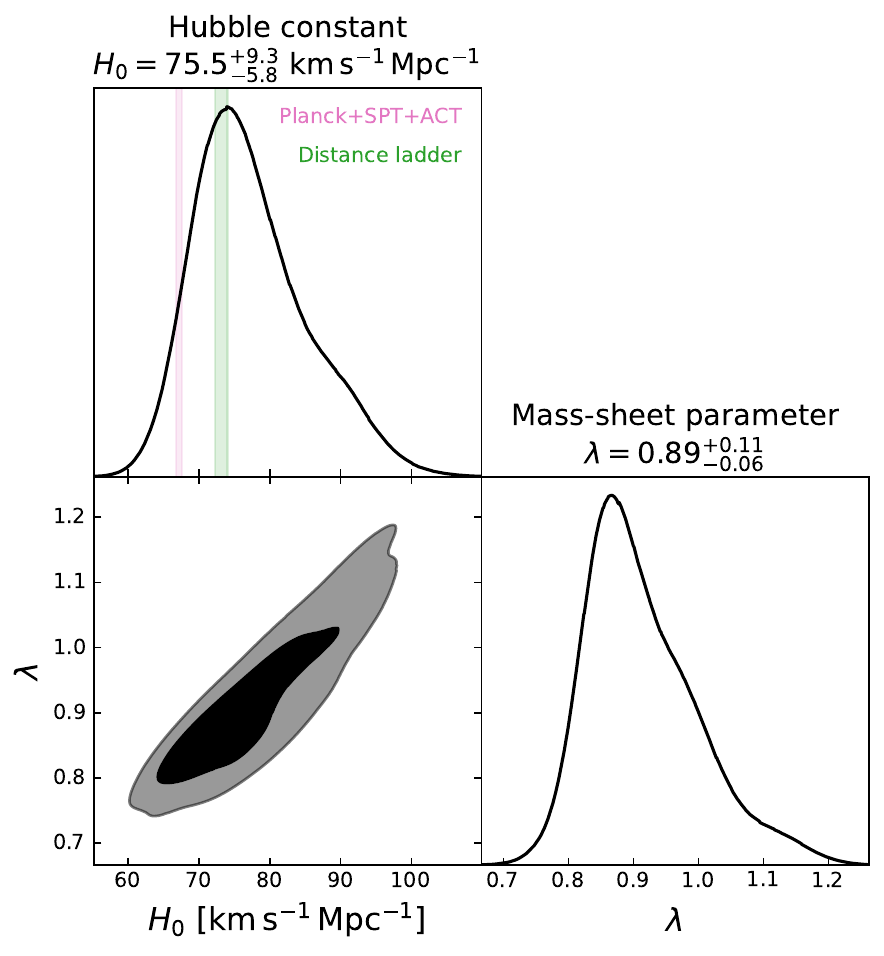}
    \caption{Joint posterior distribution on \Hc and the total MST parameter $\lambda\equiv(1-\kext)\lint$ from \fifteenthirtyseven only. For \Hc we show the results from \citet{2026PhRvD.113h3504C} and \citet{2024ApJ...973...30B} as early (Planck+SPT+ACT) and late (distance ladder) universe probes, respectively.} 
    \label{fig:cosmo_results}
\end{figure}

We infer $\Hc = 75.5^{+9.3}_{-5.8}\ {\rm km\,s^{-1}\,Mpc^{-1}}$, our primary \Hc measurement from \fifteenthirtyseven by marginalizing over \lenstro and \glee results presented in Sect.~\ref{sec:dynamical_weighting} with equal weights. We show the joint $\Hc-\lambda$ posterior distribution in Fig.~\ref{fig:cosmo_results}. We exclude from these results the spherically aligned JAM models, as they provide a measurably worse fit to the observed stellar kinematics (right panel of Fig.~\ref{fig:dynamical_model_posterior}). The total error budget on \Hc amounts to $10\%$ and incorporates both the internal and external MSD as well as differences among lens modeling codes. Separating the modeling codes we obtain $\Hc = 77.2^{+10.1}_{-6.8}\ {\rm km\,s^{-1}\,Mpc^{-1}}$ with \lenstro and $\Hc = 74.7^{+9.2}_{-5.6}\ {\rm km\,s^{-1}\,Mpc^{-1}}$ with \glee, statistically agreeing with each other.

Our \Hc measurement is in excellent agreement with other fully independent late-Universe probes of such as those from the distance ladder as indicated in Fig.~\ref{fig:cosmo_results} \citep[reference value taken from][]{2024ApJ...973...30B}. The current uncertainties also make our measurement consistent with early-Universe probes at the $\sim1\sigma$~level, for instance the combination of Planck, South Pole Telescope (SPT) and Atacama Cosmology Telescope (ACT) results \citep{2026PhRvD.113h3504C}. These methods typically combine observations either of many objects (distance ladder) or covering large areas of the sky. Our measurement is instead directly inferred from observations of a single system and remains independent of all these other methods. Improved spectroscopic and imaging data are all expected to shrink the error budget, including the systematics due to the MSD. The few percents precision of the time delays set a promising floor of what will be reachable in the future with \fifteenthirtyseven alone.

\begin{table*}[!th]
\renewcommand{\arraystretch}{1.5}
\centering
\caption{Properties of \fifteenthirtyseven, after combining the lensing constraints with stellar kinematics and LoS within a flat \lcdm cosmology. \label{tab:constraints_summary}}

\begin{tabular}{c c c c c c c}
\hline\hline
Lensing constraints & $\theta_{\rm E,G1}$\ [\arcs] & $\gamma_{\rm pl}$ & $q_{\rm m}$ & $D_{\Delta t}^{\rm model}$ [Mpc] \tablefootmark{\dag} & $\Ddt$ [Mpc] & $\lint$ \\
\hline
All & $1.378^{+0.009}_{-0.020}$ & $1.98^{+0.03}_{-0.05}$ & $0.83^{+0.01}_{-0.02}$ & $3354^{+144}_{-154}$ & $3713^{+316}_{-402}$ & $0.89^{+0.09}_{-0.07}$ \\
\lenstro & $1.360^{+0.005}_{-0.004}$ & $2.01^{+0.02}_{-0.02}$ & $0.83^{+0.02}_{-0.02}$ & $3286^{+132}_{-143}$ & $3628^{+357}_{-415}$ & $0.89^{+0.09}_{-0.07}$ \\
\glee & $1.386^{+0.002}_{-0.002}$ & $1.94^{+0.02}_{-0.03}$ & $0.83^{+0.01}_{-0.01}$ & $3418^{+107}_{-115}$ & $3751^{+308}_{-414}$ & $0.89^{+0.10}_{-0.06}$ \\
\hline
\end{tabular}

\tablefoot{For reference, the effective radius of G1 is $\reff = 1\farcs57 \pm 0\farcs05$ (measured independently of the lens models, see Sect.~\ref{app:ssec:lens_light_model}). The first row combines \lenstro and \glee results with equal weights. Only models obtained with cylindrical JAM (solid lines in Fig.~\ref{fig:dynamical_model_posterior}) are considered. The reported values correspond to the median,
16th and 84th percentiles of the posterior distributions.\\
\tablefoottext{\dag} $\Ddt^{\rm model}$ is the time-delay distance informed by strong lensing only, as opposed to \Ddt which incorporates the effects of the MSD (Eq.~\ref{eq:ddt_full_mst}).
}
\end{table*}

\section{Conclusion \label{sec:conclusion}}

We have presented the end-to-end cosmographical analysis of the quad lensed quasar \fifteenthirtyseven, with particularly precise time delays and informative multi-band HST data. Our analysis, which have been performed blindly, provides two main results. First, it provides a new, direct measurement of \Hc from an individual lensed quasar, with $\Hc = 75.5^{+9.3}_{-5.8}\ {\rm km\,s^{-1}\,Mpc^{-1}}$ in flat \lcdm cosmology with a prior on \Om from Pantheon+ (although the choice of \Om prior is subdominant). Second, it provides posterior distributions on galaxy structure and cosmological distances to inform further our TDCOSMO population-level inferences, approaching further the targeted percent-level precision on \Hc from time-delay cosmography.

Beyond our main cosmographical measurement, we have presented the following results:
\begin{enumerate}
    \item We incorporated some of the most precise time delays in a lensed quasar to date ($\sim2\%$ precision) in lens models constrained from multi-band HST imaging data using two independent modeling codes ran by independent teams;
    \item We measured resolved stellar kinematics of the main deflector from VLT/MUSE observations, revealing a clear rotation field and providing velocity dispersions that further constrain the deflector's mass distribution;
    \item We characterized in details the LoS and environment of the lens, both spectroscopically and photometrically, to incorporate explicitly (via perturbers) and statistically (via the external convergence) the effect of additional structures along the LoS;
    \item We combined strong lensing (imaging and time delays), resolved stellar kinematics and LoS information to explicitly incorporate the effects of the MSD, the leading systematic error term in time-delay cosmography with single systems;
    \item Within a flat \lcdm model and \Om informed by observations of Type Ia supernovae, our combined constraints result in a $10\%$ measurement of \Hc from a single system, without any systematics shared with other \Hc probes.
\end{enumerate}

The total uncertainty of \Hc may be perceived as disappointing, given the promises of the precise time delays of \fifteenthirtyseven. However, our inference conservatively incorporates all known sources of systematic errors, turning the MSD into a statistical uncertainty and including fundamental differences between lens modeling methods \citep{2024A&A...692A..87G}. Importantly, this first end-to-end analysis of \fifteenthirtyseven prepares the ground for significant improvements thanks to incoming new data sets. At the time of writing these lines, the JWST started gathering observations of this system. Resolved spectroscopic from JWST/NIRSpec will vastly surpass the VLT/MUSE used here, significantly improving pixel scale, spatial and spectral resolutions (see Knabel et al. in prep., Mozumdar et al. in prep. for comparable data). Similarly, near-infrared imaging data from JWST/NIRCam will provide similar resolution as the bluer HST filters used here but with much higher S/N and more extended arcs \citep[][Williams et al. in prep.]{2025A&A...703A.118W,2026arXiv260708576L}. Within the same modeling and inference framework, we expect these new observations to provide more constraints on the gravitational potential of the main deflector both from strong lensing and stellar dynamics, and ultimately improve the precision on \Hc from \fifteenthirtyseven only.

Beyond a new single-system measurement, we increase the size of our sample of time-delay lensed quasars with added lensing constraints, a spatially resolved kinematics map and LoS information. Future TDCOSMO milestone results (TDCOSMO Collaboration 2026, in prep.) will incorporate these new products to further inform the properties of the deflector galaxy population jointly with the expansion rate in a variety of cosmological models.


\begin{acknowledgements}
We thank all members of the TDCOSMO collaboration for crucial help and feedback that improved this analysis and manuscript. We also thank Raoul Cañameras for helping with the characterization of galaxy G5, and Riccardo Seppi for helping with the galaxy group characterization.
AG acknowledges support from the Swiss National Science Foundation (SNSF) through the PostDoc Mobility Return Grant P5R5-2\_235360.
SE and SHS thank the Max Planck Society for support through the Max Planck Fellowship for SHS. This work is supported in part by the Deutsche Forschungsgemeinschaft (DFG, German Research Foundation) under Germany's Excellence Strategy -- EXC-2094 -- 390783311. 
DJ acknowledges support by the First Rand Foundation, South Africa, and the Centre National de la Recherche Scientifique of France.
MM acknowledges support by the SNSF through Ambizione grant PZ00P2\_223738.
We acknowledge support by NSF through grants NSF-AST-2407277, and NSF-AST-1836016, and from the Moore Foundation through grant 8548.
TA acknowledges support from ANID-FONDECYT Regular Project 1240105 and the ANID BASAL project FB210003.
This work has made use of CosmoHub, developed by PIC (maintained by IFAE and CIEMAT) in collaboration with ICE-CSIC. It received funding from the Spanish government (grant EQC2021-007479-P funded by MCIN/AEI/10.13039/501100011033), the EU NextGeneration/PRTR (PRTR-C17.I1), and the Generalitat de Catalunya.
VM acknowledges support from ANID FONDECYT Regular grant number 1231418 and Centro de Astrof\'{\i}sica de Valpara\'{\i}so.
KCW is supported by JSPS KAKENHI Grant Numbers JP24K07089, JP24H00221.
DE acknowledges support from the SNSF through grant agreement \#200021\_212576.
This research has made use of \textsc{SciPy} \citep{Virtanen2020scipy}, \textsc{NumPy} \citep{Oliphant2006numpy,VanDerWalt2011numpy}, \textsc{Matplotlib} \citep{Hunter2007matplotlib}, \textsc{Astropy} \citep{astropy2013,astropy2018}, \textsc{Jupyter} \citep{Kluyver2016jupyter} and \textsc{GetDist} \citep{Lewis2019getdist}.

\end{acknowledgements}

\bibliographystyle{aa}
\bibliography{biblio}


\begin{appendix}

\section{Identification of perturbers \label{app:sec:perturbers}}

\subsection{Objects catalog \label{app:ssec:catalog_making}}

\begin{figure*}[h!]
\sidecaption
    \includegraphics[width=\linewidth]{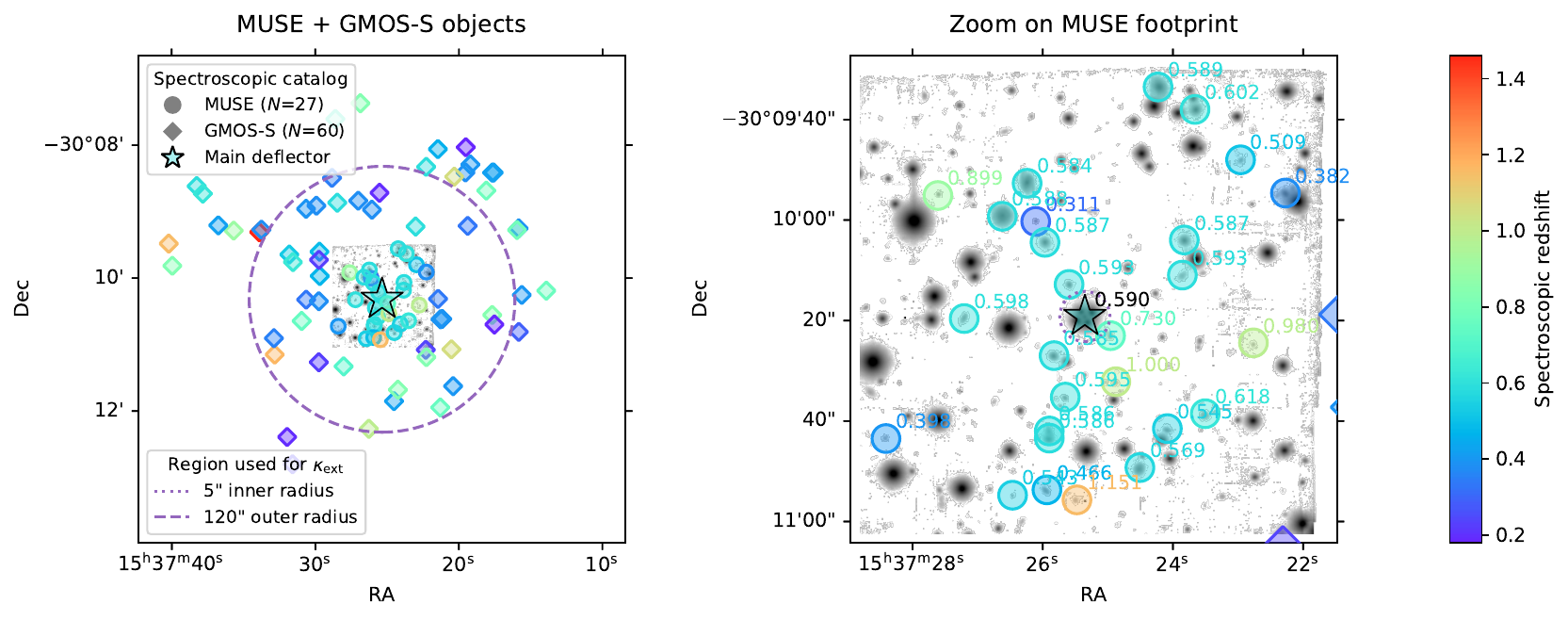}
    \caption{Environment of \fifteenthirtyseven showing the 87 objects for which we measured spectroscopic redshifts (excluding stars and low S/N objects, see Sect.~\ref{app:ssec:catalog_making} for details). The circles with radius $5\arcs$ and $120\arcs$ indicates the extent of the region considered to infer $\kext$ (see Sect.~\ref{ssec:los_kappa_ext}).}
    \label{app:fig:environment}
\end{figure*}

Spectra of the brightest galaxies (down to $i \sim 23$ mag) in the field of view up to a projected distance of $\sim 3$\,arcmin from the lens, have been obtained using the MUSE data which covers the central $90\times90$\arcsec region (Sect.~\ref{ssec:lens_kinematics}), and through dedicated multi-object spectroscopic observation with GMOS-S (PI: Buckley-Geer; PID: GS-2022A-Q-135). MUSE sources were detected with \textsc{SExtractor} \citep{BertinArnouts1996} on the white-light image, with spectra extracted from the resulting segmentation map and redshifts measured via template cross-correlation with MARZ \citep{Hinton2016}. An independent MUSELET emission-line search \citep{Piqueras2019} found no additional sources; a final visual quality check yielded the MUSE catalog of 124 sources (57 stars, 27 secure redshifts). The remaining targets had only a single emission line or were too noisy to enable a solid redshift estimate.

The GMOS-S observations used four masks for a total of 136 targets. The target selection for the masks was done using GMOS-S $i-$band imaging data that was obtained as part of the observation program. We did not include targets that had already been observed using MUSE. The Gemini data were processed, the spectra extracted and the redshifts measured as described in \citet{BuckleyGeer2020} with one difference. Instead of using the Gemini IRAF \footnote{\url{https://nsf-noirlab.gitlab.io/csdc/usngo/gmos-cookbook/}} \texttt{gswavelength} task, we developed our own wavelength calibration software based on the RASCAL package~\footnote{\url{https://github.com/jveitchmichaelis/rascal}} \citep{2020ASPC..527..627V}. We obtained 60 secure redshifts. The MUSE and GMOS objects were then combined into one catalog totaling 87 objects. We show all detected objects and a zoom on the MUSE footprint in Fig.~\ref{app:fig:environment}.

\subsection{Individual galaxies \label{app:ssec:perturbers_identification}}

As in \citet{BuckleyGeer2020}, we mostly follow the procedure described in \citet{2019MNRAS.490..613S} to characterize the environment of \fifteenthirtyseven. We first build a catalog of objects from merging detections from MUSE, which covers the inner $90\arcs$ box around the lens, with GMOS-S that expands it about $3'$ (see Sect.~\ref{app:ssec:catalog_making} for details). Then, for each detected object classified as a galaxy---considered as perturber candidates---, we perform spectral energy distribution (SED) fitting of $\textit{griz}$ filters to estimate stellar masses. We then use the $M-\sigma$ scaling relation from \citet{Auger2011} to obtain the central velocity dispersion, from which we obtain an approximate Einstein radius assuming a singular isothermal sphere (SIS, see Sect.~\ref{app:ssec:mass_profiles} for a definition). The Einstein radius of each perturber candidate $\theta_{\rm E, pert}$ is then used to compute the flexion shift $\flexshift$ \citep{McCully2017} as
\begin{align}
\label{eq:def_flexion_shift}
    \Delta_{3}x(\theta_{\rm E, gal}) = f(\xi)\frac{\theta^2_{\rm E, main}\theta^2_{\rm E, pert}}{\delta_{\rm main}^3} \ ,
\end{align}
where $\theta_{\rm E, main}$ is the Einstein radius of the main deflector and $\delta_{\rm main}$ is the angular separation between the main deflector and the perturber. The scaling factor $f(\xi)$ is a piecewise function defined as
\begin{equation}
    f(\xi) = \begin{cases}
        1 & \text{if } D_{\rm p} < D_{\rm d} \ {\rm (background\ perturber)}\ , \\
        (1-\xi)^2  & \text{otherwise} \ {\rm (foreground\ perturber)}\ , 
\end{cases} \end{equation}
evaluated at the distance ratio
\begin{align}
    \xi \equiv \frac{D_{\rm dp}D_{\rm s}}{D_{\rm p}D_{\rm ds}} \ , 
\end{align}
where $D_{\rm dp}$ and $D_{\rm p}$ are the angular diameter distances between the main deflector and the perturber, and between the observer and perturber, respectively.

We use $\flexshift$ as a criterion to select galaxies and galaxy groups whose projected distance to the main deflector and lensing strength may bias \Hc at the percent level if they are not explicitly included in the lens models (see also Sect.~\ref{sec:lens_modeling}). As for past systems, we follow \citet{2019MNRAS.490..613S} and select galaxies and groups of galaxies that have a flexion shift $\log_{10}\flexshift > -4$, within their $1\sigma$ uncertainties. This criterion selects the four objects labeled G2 to G5 in Fig.~\ref{fig:color_image}. However, noticing that G5 is an emission-line galaxy, we expect the procedure of \citet{BuckleyGeer2020} to be less accurate as it is better suited for elliptical galaxies \citep[in particular the scaling relation from][]{Auger2011}. Therefore, we re-estimate the flexion shift using the stellar-host for star-forming galaxies of \citet{2019MNRAS.488.3143B} and find that it does not pass the flexion shift threshold anymore (see Sect.~\ref{ssec:env_perturbers}). We list the main properties of the perturbers in Table~\ref{app:tab:perturbers_props}.

\subsection{Galaxy groups \label{app:ssec:groups_identification}}

Following a similar procedure we also identified two galaxy-group candidates at redshifts $z\approx0.39$ and $z\approx0.59$. With $\log \Delta_3x=-2.76^{+0.81}_{-1.08}$, only the $z\approx0.59$ group passes the flexion shift criterion. However, upon inspection of the spectroscopic catalog, we noticed one galaxy, located north west from the lens at $z\approx0.59$, was missing. This galaxy is among the three brightest objects around this redshift, therefore an additional group member candidate that would drive the group mass and centroid estimates. After adding the galaxy, we ran the friend-of-friend (FoF) group-finding algorithm of \citet{2014A&A...566A...1T}. Testing plausible ranges of FoF linking parameters \citep{2017A&A...602A.100T} around the candidate group yields 10-18 members (vs. 8 initially), a centroid $17$-$28''$ from the lens (vs. $12''$), and a velocity dispersion of $530$-$1420\ {\rm km\,s}^{-1}$ (vs. $280\ {\rm km\,s}^{-1}$). These results point to significant uncertainty in the group characterization as they appear unrealistically large for a group and the X-ray upper mass we estimate from X-ray observations.


The available \emph{Chandra} observations (PI: Pooley; PID: 22700509) do not show evidence for extended X-ray emission on top of the emission from the lensed quasar. We follow the procedure outlined in \citet{LHAASO23} to extract an upper limit to the X-ray emission of the group and turn it into an upper limit on the group's mass. Namely, we merged the three available \emph{Chandra} exposures (observation IDs 22019, 23828, and 26272) and extracted \emph{Chandra}/ACIS-I count images and exposure maps in the [0.5-2] keV band from the merged event list. We then extracted the surface brightness profile centered on the quasar, and searched for potential X-ray emission in its surroundings compared to the background level. Assuming an intrinsic source described by an APEC model \citep{APEC} with a temperature of 2 keV, the non-detection of X-ray emission over the background level places an upper limit of $1.8\times10^{43}$ erg/s to the luminosity of the group in the [0.5-2] keV. Converting the luminosity into an upper limit on the halo mass through the $L_X-M$ relation of \citet{Lovisari:2020}, we set an upper limit of $M_{500c}<4\times10^{13}\ {\rm M}_\odot$ on the mass of this potential group. We also searched for cluster catalogs from large surveys that may contain the candidate group; however, given the location of proximity of \fifteenthirtyseven to the galactic plane, this region is excluded from clustering analyses to avoid contamination by dust and stellar fields. With the data in hand, we are therefore unable to robustly characterize the mass and position of the candidate group at $z\sim0.59$.

Our lens models explicitly include four of the group member candidates (G1 to G4). The orientation of the external shear already included in all our lens models qualitatively matches a group-scale halo with centroid located northward of the main deflector. Moreover, we did not exclude other group member candidates from the catalog used for inferring the external convergence, hence first-order effects due to an overdensity of galaxies in the vicinity of the lens are also included in our analysis.

\begin{table*}[ht]
    \caption{Properties of galaxies identified as perturbers (labeled in Fig.~\ref{fig:color_image} and see Sections \ref{ssec:env_perturbers} and \ref{app:sec:perturbers} for details).}
	\renewcommand{\arraystretch}{1.6}
    \centering
    \begin{tabular}{ccccccccc}
        \hline\hline
        Label & $z$ & RA [\,$^\circ$\,] & Dec [\,$^\circ$\,] & $\delta_{\rm G1} x\ [\,\arcs\,]$ & $\delta_{\rm G1} y\ [\,\arcs\,]$ & $\theta_{\rm E, SIS}\ [\,\arcs\,]$ & $\sigma_{v,\rm los}$ [km\,s$^{-1}$] & $\log_{10}(\Delta_3 x)$ \\
        \hline
        G2 & $0.593$ & $234.356650$ & $-30.169576$ & $-3.0949$ & $6.3900$ & $0.23^{+0.04}_{-0.03}$ & $121.7^{+9.4}_{-9.4}$ & $-3.54^{+0.13}_{-0.14}$ \\
        G3 & $0.585$ & $234.357609$ & $-30.173515$ & $-6.1571$ & $-7.7610$ & $0.24^{+0.03}_{-0.03}$ & $123.3^{+7.7}_{-7.7}$ & $-3.94^{+0.11}_{-0.11}$ \\
        G4 & $0.587$ & $234.358184$ & $-30.167250$ & $-7.8932$ & $14.8025$ & $0.32^{+0.07}_{-0.06}$ & $142.0^{+15.2}_{-15.2}$ & $-4.38^{+0.18}_{-0.20}$ \\
        G5 & $0.730$ & $234.353989$ & $-30.172410$ & $5.1158$ & $-3.8012$ & $0.13^{+0.03}_{-0.03}$ & $98.2^{+8.5}_{-8.5}$ & $-4.22^{+0.20}_{-0.22}$ \\
        \hline
    \end{tabular}
	\label{app:tab:perturbers_props}
    \tablefoot{From left to right: label, redshift ($z$ precise up to the last quoted digit), RA and Dec coordinates, relative positions with respect to main deflector G1 (along the HST/F160W image axes), Einstein radius assuming an SIS mass distribution, LoS velocity dispersion and flexion shift (Eq.~\ref{eq:def_flexion_shift}) computed from the reported Einstein radius values.\\
    \tablefoottext{$\dag$}{For G2, G3 and G4, the velocity dispersion is measured from the MUSE data, while for G5 it is inferred from its estimated halo mass assuming an SIS total mass distribution (see Sect.~\ref{ssec:env_perturbers} for details).}
    }
\end{table*}

\section{Kinematics extraction on quasar-subtracted data \label{app:sec:kinem_qsosub}}

\begin{figure}
    \centering
    \includegraphics[width=\linewidth]{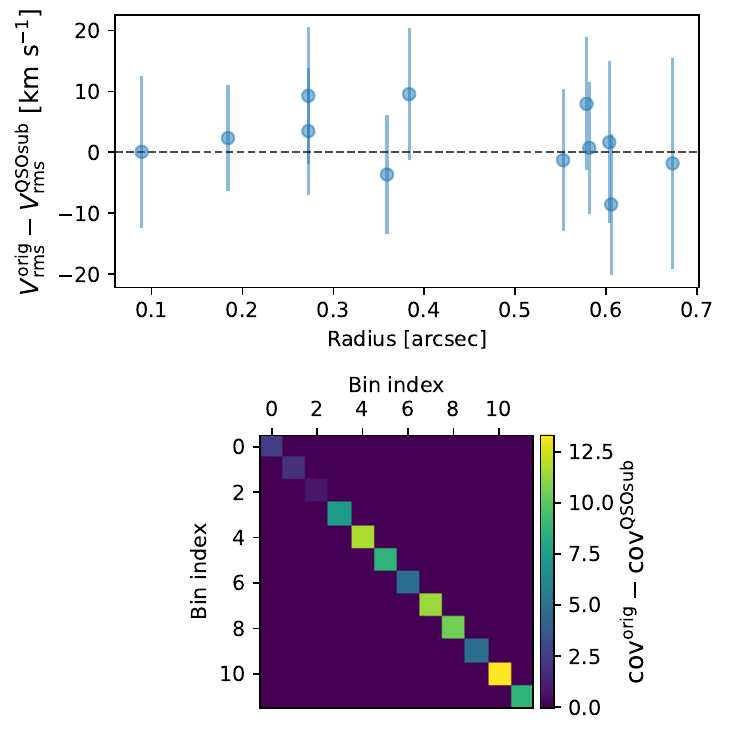}
    \caption{Difference between $V_{\rm rms}$ measurements extracted on the original MUSE cube (``orig'') and the quasar-subtracted cube (``QSOsub''). The top panel shows the difference in radial profiles (with errorbars from the ``orig'' extraction), while the bottom panel shows the difference in covariance matrices. Both the best-fit values and the covariances are consistent within the uncertainties.}
    \label{app:fig:v_rms_comp_qsosub}
\end{figure}

Our baseline stellar kinematics measurements are performed on the original MUSE datacube. As an additional systematic test, we performed the same measurements on the quasar-subtracted datacube. We still keep the fixed quasar templates, although we find that they get zero weights attribute, as expected. The comparison between the two alternative measurements are shown in Fig.~\ref{app:fig:v_rms_comp_qsosub}. We find that the two velocity dispersion maps remain statistically consistent with each other. The total uncertainties (diagonal of the covariance matrix) are slightly larger in our baseline measurements, by $\sim 4\ {km\,s}^{-1}$ at maximum. Therefore, we conclude that the choice of data used for fitting spectral templates has negligible impact on the kinematics maps.

\section{Lens modeling details \label{app:sec:lens_model_details}}

\subsection{Lens modeling with \lenstro \label{app:ssec:lenstronomy_modeling}}

\subsubsection{Lens model setup}

The lens model likelihood has three main terms (in addition to prior terms associated to model parameters), all assuming Gaussian distributions. These terms are the imaging term conditioned on the HST data and associated uncertainties, the time-delay term conditioned on the measured time delays and covariance matrix, and an source-plane position term to ensure the quasar images fulfill the lens equation (Eq.~\ref{eq:lens_equation}). For the source-plane position term, we follow standard assumptions from past \lenstro analyses and assume a Gaussian prior with $0.5$ mas standard deviation on the source-plane quasar position (computed as the mean position over the four quasar images), and penalize models whose individually ray-traced image deviate by more than $1$ mas from the source-plane quasar position. These assumptions are consistent with the results of \citet{BirrerTreu2019}, from which we find that a precision of 9 mas is required in the source position to match the most precise time delay precision \timedelayat{A}{C}. Given the magnification and the multiplicity of the images, this is matched by an astrometric precision of about 26 mas in the image plane per image, under fixed lens model. From our HST imaging data we are able to perform a reliable astrometry below 6 mas \citep{Schmidt2023} and thus we expect any unaccounted astrometric uncertainty to be subdominant in our error budget.

We use two concentric Sérsic profiles to model the main deflector surface brightness, fixing the Sérsic index of each profile to 1 (exponential) and 4 (de~Vaucouleurs) to avoid degeneracies. We also consider two concentric chameleon profiles (Eq.~\ref{eq:def_chameleon}) as a variation of the baseline model, since its different parametrization changes the high-dimensional likelihood surface allowing the model to potentially converge to different local minima.

We use an elliptical Sérsic profile (Eq.~\ref{eq:def_sersic_ellipse}) combined with shapelets \citep{Birrer2015} with joint centroids to model the quasar host galaxy. We use this pair of profiles for each band, although we join some of the parameters between the band, similar to \citet{Shajib2020}. Specifically, the Sérsic index, ellipticity parameters and shapelet scale are shared between the UVIS bands. As the position of the quasar in source plane is mostly constrained by the multiple point source images, we follow \citet{Birrer2019,Shajib2020} and fix the centroid of the extended source --- the profiles mentioned above --- to the optimized position of the quasar in source plane. While this approach is likely to be justified for an isolated and symmetric host galaxy, we notice that the slight apparent asymmetry (see rightmost column in Fig.~\ref{fig:bestfit_lens_models}) of the extended source light could hint towards an off-centered quasar with respect to its host. For this reason, we ran additional models based on our baseline power-law setup, dropping this assumption. The time delay distance posterior distributions, after marginalization combination according to the BIC (see Sect.~\ref{ssec:bic_weighting} for details), is affected by less than a percent. Hence, we conclude that joining the quasar position in source plane with the center of the host surface brightness does not significantly bias our model, as shapelets are sufficiently flexible to account for this offset.

To mitigate numerical inaccuracies from centering and evaluating light profiles and deflection angles on a finite grid of pixels, we subsample our model evaluations in certain regions of the data. These regions contain pixels close to the lens galaxy centroid, quasar images and the brightest parts of the arc. Inside those regions, we evaluate the model at three times the data resolution. We also subsample the model of the PSF to three times the data resolution, to prevent potential sources of biases in lens model parameters \citep{2022A&A...667A.123S,2024A&A...692A..87G}.

\subsubsection{Modeling sequence}

We build an initial PSF by median-stacking field stars per filter, enforced to $90^\circ$-rotational symmetry, with a per-pixel error map from the star-cutout variance. This is iteratively refined between PSO runs following the standard procedure of \citep[as described in][]{Birrer2019,Shajib2020,Chen2016}, and further validated on the Time Delay Lens Modeling Challenge (TDLMC) \citep{Ding2021_TDLMC}, and the resulting median PSF is fixed for the final models.

Given the $\sim50$-dimensional parameter space, we use a sequence of PSO runs, progressively releasing parameters, before a final MCMC. After each PSO, we refine residual offsets between the HST filters, mainly driven by the position of quasar images. The slope of the power-law profile is initially fixed to isothermal and released during the last PSO. The reference scale $\beta_{\rm sh}$ for each shapelets sets is set to the effective radius of the Sérsic profile in each filter. The half-light radius is a good initial guess for $\beta_{\rm sh}$ as it represents an average scale of the surface brightness. Since to first-order, the morphology of the quasar host galaxy is well described by the elliptical Sérsic profile, we enforce the shapelets to only target small-scale deviations to the Sérsic profile. We thus ensure that the Sérsic profile does not vary excessively after the addition of shapelets by introducing Gaussian priors on the Sérsic parameters centered on their values fitted without shapelets, with a 10\% standard deviation. This strategy has also been validated on the TDLMC by the EPFL team \citep[][]{Ding2021_TDLMC}. After a last PSO, we run a MCMC to sample the full joint posterior distribution. On 72 CPU cores (Intel Xeon IceLake-SP), one \lenstro model variation takes on average one full day to run, including the PSO, MCMC and postprocessing of the chains.

\subsubsection{Lens model variations \label{app:sssec:model_variations}}

We run several variations of the \lenstro model, and combine using the BIC as described in Sect.~\ref{ssec:bic_weighting}. We consider the following model variations:
\begin{itemize}
    \item maximum order of the source shapelets set in each filter $\left\{\nmax^{\ir},\, \nmax^{\uvisone},\, \nmax^{\uvistwo}\right\}$: $\{8,\, 5,\, 7\}$, $\{9,\, 6,\, 8\}$ and $\{10,\, 7,\, 9\}$;
    \item radius of main mask region centered on the lens: 4\farcs1 and 4\farcs3;
    \item main deflector light profile in the \ir filter: double-Sérsic and double-chameleon profile (while keeping double-Sérsic for \uvisone and \uvistwo filters);
\end{itemize}

We calculate the BIC using our fiducial $4\farcs1$ mask and take the sample that has the largest log-probability from the MCMC chain. To estimate the BIC uncertainty $\sigma_{\Delta \rm BIC}$, we consider two terms. The first term  $\sigma^{\rm num}_{\Delta \rm BIC}$ is of numerical origin, and pertains to the inherent stochasticity of the optimizations and sampling step. We estimate  $\sigma^{\rm num}_{\Delta \rm BIC}$ by running each model variation twice, changing only the random seed to initialize the PSO and MCMC, and set $\sigma^{\rm num}_{\Delta \rm BIC}$ to the mean BIC difference between such identical runs. The second BIC uncertainty term, $\sigma^{\rm model}_{\Delta \rm BIC}$, takes into account the sparsity with which we sample the space of all possible model variations. We set $\sigma^{\rm model}_{\Delta \rm BIC}$ to the scatter among pairs of models that only differ by one setting (e.g., different shapelets \nmax, lens light profile, or likelihood mask size). We take the quadratic sum of both uncertainty terms to obtain $\sigma_{\rm \Delta BIC}$ to use Eq.~\ref{eq:bic_weights}.

\subsubsection{\lenstro lens modeling results}

\begin{table*}
    \caption{Model variations performed with \lenstro and their resulting BIC weights.}
    \label{app:tab:lenstro_model_variations}
    \renewcommand{\arraystretch}{1.2}
    \centering
    \begin{tabular}{lccc}
        \hline\hline
        Model variation & BIC & $\Delta$BIC & $w_{\rm BIC}^\ast$ (Eq.~\ref{eq:bic_weights})  \\
        \hline
        Sérsic, G2--G4, $n_{\rm max}=\{10,7,9\}$, mask $r<4\farcs1$ & 78260 & 0 & 0.1737 \\
        Sérsic, G2--G4, $n_{\rm max}=\{9,6,8\}$, mask $r<4\farcs1$ & 78268 & 8 & 0.1723 \\
        Sérsic, G2--G5, $n_{\rm max}=\{9,6,8\}$, mask $r<4\farcs1$ & 78406 & 146 & 0.1480 \\
        Sérsic, G2--G4, $n_{\rm max}=\{9,6,8\}$, mask $r<4\farcs3$ & 78512 & 252 & 0.1299 \\
        Chameleon, G2--G4, $n_{\rm max}=\{9,6,8\}$, mask $r<4\farcs1$ & 78540 & 280 & 0.1253 \\
        Chameleon, G2--G4, $n_{\rm max}=\{8,5,7\}$, mask $r<4\farcs1$ & 78853 & 593 & 0.0782 \\
        Sérsic, G2--G5, $n_{\rm max}=\{10,7,9\}$, mask $r<4\farcs1$ & 78887 & 627 & 0.0737 \\
        Sérsic, G2--G4, $n_{\rm max}=\{10,7,9\}$, mask $r<4\farcs3$ & 79020 & 760 & 0.0578 \\
        Chameleon, G2--G4, $n_{\rm max}=\{10,7,9\}$, mask $r<4\farcs1$ & 79445 & 1185 & 0.0228 \\
        Sérsic, G2--G4, $n_{\rm max}=\{8,5,7\}$, mask $r<4\farcs3$ & 79615 & 1355 & 0.0146 \\
        Sérsic, G2--G4, $n_{\rm max}=\{8,5,7\}$, mask $r<4\farcs1$ & 80103 & 1843 & 0.0033 \\
        Sérsic, G2--G5, $n_{\rm max}=\{8,5,7\}$, mask $r<4\farcs1$ & 80657 & 2397 & 0.0004 \\
        \hline
    \end{tabular}
    \tablefoot{The uncertainty on the $\Delta {\rm BIC}$ is $\sigma_{\Delta \rm BIC}=785$ and all $\Delta{\rm BIC}$ values should be interpreted within this uncertainty. The model variation labels specify the lens light profile (Sérsic or Chameleon), the range of perturbers included (G2--G4 or G2--G5), the shapelets maximal order \nmax for each band $\{\ir,\uvisone,\uvistwo\}$ and the outer radius of the likelihood mask ($4\farcs1$ or $4\farcs3$). Each model variation was performed twice, although for conciseness we only show a single realization in the table (see Sect.~\ref{app:sssec:model_variations} for details).}
\end{table*}

We show in Fig.~\ref{fig:bestfit_lens_models} the \lenstro lens model that has the preferred BIC value. We also summarize the model variations and their corresponding BIC weights in Table~\ref{app:tab:lenstro_model_variations}. We estimate $\sigma_{\Delta \rm BIC}^{\rm num}=558$ and $\sigma_{\Delta \rm BIC}^{\rm model}=552$, leading to an uncertainty in $\Delta$BIC of $\sigma_{\Delta \rm BIC}=785$. After BIC weighting of each model variation, we find obtain the posterior distributions shown in Fig.~\ref{fig:lens_model_posterior}.

\subsection{Lens modeling with \glee \label{app:ssec:glee_modeling}}

\subsubsection{Lens model setup}

To model the surface brightness of the main deflector, we use a combination of two S\'ersic profiles. This double-S\'ersic model is required to adequately fit the deflector light across all three filters. Both profiles share a single fixed centroid across the three bands, and we allow their S\'ersic indices to vary freely during optimization. Despite imposing no additional priors, we find that the models converge readily.

We model the surface brightness of the extended quasar host galaxy using a regularized pixel grid, while the quasar is modeled as a point source whose source-plane position is primarily constrained by the multiple lensed images. The amplitudes of the lensed quasar images are allowed to vary freely to account for possible flux ratio anomalies due to millilensing and microlensing. The grid-based reconstruction of the extended source naturally accommodates complex morphologies, inherently accounting for any potential asymmetry or physical offset between the active galactic nucleus and the host emission. Because the image cutouts do not share the exact same footprint across filters, the surface brightness modeling is initially performed in the native coordinate space of each band's cutout. We subsequently use the inferred quasar positions to align the coordinate systems across all filters for consistent surface brightness modeling. The baseline grid resolution for the extended source is $35 \times 35$ pixels in the  F160W and $55 \times 55$ in the F814W and F475X.

\subsubsection{Modeling sequence}

We model the PSF directly from the science images by stacking several non-saturated stars within the field of view. Cutouts of these stars are extracted and shifted using sub-pixel interpolation to properly center the profile, ensuring the flux in the central $3 \times 3$ (native) pixel region is symmetric to within 10\%. The recentered stars are then normalized, stacked pixel-wise using a median, and re-normalized so that the total pixel sum equals unity. To optimize the modeling process, we generate two versions of this empirical PSF. The first retains the full diffraction spikes and is used directly to model the quasar point sources. The second is cropped at the first diffraction minimum of the Airy disc to serve as the convolution kernel for the extended source and deflector light; this significantly reduces computational overhead. Finally, the PSF of the F160W band is three-times subsampled to prevent numerical inaccuracies during the modeling process (similar to the \lenstro model). The UVIS bands were not subsampled because they have twice the pixel resolution (0$\arcsec$.04/pixel) of the F160W (0$\arcsec$.08/pixel), and to reduce computational overhead. During the lens light modeling, we further refine this empirical PSF to account for subtle mismatches. We perform an iterative correction procedure, as described by \cite{Wong2017}, updating the PSF using the residuals at the locations of the lensed quasar images.

The non-linear parameter space of our lens model is high-dimensional, necessitating a robust strategy to locate the global minimum and appropriately sample the posterior distribution. We employ an iterative approach that alternates between MCMC sampling and simulated annealing (siman). We begin by estimating an initial sampling covariance matrix to guide the sampler; we do this using the highly parallelizable EMCEE \cite{Foreman-Mackey_2013}. Subsequently, we run an MCMC phase followed by simulated annealing, updating the sampling covariance matrix at each cycle to robustly explore the parameter space. This iterative sequence continues until the MCMC chains reach convergence, which we assess by analyzing their power spectra, as described in \cite{Dunkley05}. Once convergence is achieved, we run a final, extensive MCMC "production" chain to fully sample the joint posterior distribution and derive our final parameter estimates and uncertainties.

We adopt a sequential optimization strategy to robustly constrain the components of the lens system. Initially, we iteratively optimize the surface brightness parameters of the S\'ersic profiles, along with the QSO positions and amplitudes. Following a spatial realignment phase, we fix the deflector light parameters and allow the mass model parameters to vary freely. During this stage, the PSF amplitudes are deliberately reduced and re-optimized to accurately account for the blended flux of the central quasar and the underlying host galaxy. This initial mass model is constrained exclusively by the observed positions of the multiple lensed images, with an estimated error of 0\arcsec.004, which accounts for the typical astrometric precision achievable when extracting quasar point-source positions, as shown in \cite{Ertl23}. Once this preliminary mass distribution is established, we proceed with the full grid-based reconstruction of the extended source. We run an initial reconstruction cycle until convergence, followed by a second cycle in which the uncertainties in the region and the immediate vicinity of the quasar images are artificially increased (henceforth, boosted) so that the normalized residuals in the region have values of $\sigma_i \ll 1$. This step ensures that the final model is not biased by potential mismatches in the PSF which peak at the vicinity of the quasar images. Additionally, in this final step we let the Einstein radius of the explicitly modeled perturbers vary, with a Gaussian prior centered around the median and its corresponding width described in Table \ref{app:tab:perturbers_props}.

\subsubsection{Lens model variations}
To investigate potential systematics, we run several model variations. These include:
\begin{itemize}
    \item $\chi^2$ of the time delays measured at the observed image positions,
    \item thickening the arc mask by 1 pixel,
    \item boosting the QSO regions by a very high constant value (effectively giving 0 weight to the boosted region),
\end{itemize}
Each of these variations was run at two different combinations of source-grid resolution: (1) $35\times35$ in F160W paired with $55\times55$ in both UVIS bands, and (2) $45\times45$ in F160W paired with $65\times65$ in both UVIS bands.

Prior to applying the BIC weighting, we estimate the associated uncertainty $\sigma_{\rm \Delta BIC}$. This is achieved by perturbing the baseline grid resolution independently in each band, using 1-pixel increments over a range of $\pm 5$ pixels with respect to the baseline, a procedure described in \cite{2022A&A...667A.123S}. This results in $\sigma_{\rm \Delta BIC}=683$. We then compute the BIC weights for each chain using Eq.~\ref{eq:bic_weights}, the resulting weights are shown in table \ref{app:tab:glee_model_variations}.

\begin{table*}
    \caption{Model variations performed with \glee and their resulting BIC weights.}
    \label{app:tab:glee_model_variations}
    \renewcommand{\arraystretch}{1.2}
    \centering
    \begin{tabular}{lccc}
        \hline\hline
        Model variation & BIC & $\Delta$BIC & $w_{\rm BIC}^\ast$ (Eq.~\ref{eq:bic_weights}) \\
        \hline
        \multicolumn{4}{l}{\textit{Combination B: $45\times45$ (F160W) $+$ $65\times65$ (UVIS)}} \\
        Fiducial                                             & 10651 & 0    & 0.3350 \\
        $\chi^2$ of time delays at observed image positions  & 10668 & 17   & 0.3284 \\
        Thickened arc mask by 1 pixel                        & 10838 & 187  & 0.2629 \\
        Flat high constant boost                             & 11502 & 851  & 0.0715 \\
        \hline
        \multicolumn{4}{l}{\textit{Combination A: $35\times35$ (F160W) $+$ $55\times55$ (UVIS)}} \\
        Fiducial                                             & 12745 & 2094 & 0.0007 \\
        Flat high constant boost                             & 12749 & 2098 & 0.0007 \\
        $\chi^2$ of time delays at observed image positions  & 12783 & 2132 & 0.0006 \\
        Thickened arc mask by 1 pixel                        & 13091 & 2440 & 0.0001 \\
        \hline
    \end{tabular}
    \tablefoot{The uncertainty on the $\Delta {\rm BIC}$ is estimated at $\sigma_{\Delta \rm BIC}=683$ and all $\Delta{\rm BIC}$ values should be interpreted within this uncertainty. Similar to Table \ref{app:tab:lenstro_model_variations} but for the \glee model. The model variations represent different systematic tests to investigate potential biases, each performed at two source-grid resolution combinations.}
\end{table*}

\subsubsection{\glee lens modeling results}

We showed in the bottom part of Fig.~\ref{fig:bestfit_lens_models} the \glee lens model with the best BIC. The \glee lens model, informed by the time delays, assumes a uniform prior on the Hubble constant directly (denoted $\Hc^{\rm model}$) while \lenstro assumes a uniform prior on $\Ddt^{\rm model}$. We correct for this mismatch for consistency between the two codes, choosing to keep a uniform prior on $\Ddt^{\rm model}$ as the reference, which allows us to include our \fifteenthirtyseven constraints in future TDCOSMO hierarchical inferences. In order to convert the original \glee joint posterior distributions to one that encodes a uniform prior on $\Ddt^{\rm model}$, we re-weight the MCMC samples with weights (see Williams et al. in prep. for the full derivation):
\begin{align}
\label{eq:weights_prior_ddt_hc}
    w_{\rm prior} = \frac{\Ddt^{\rm model}}{\Hc^{\rm model}} \ .
\end{align}

\subsection{Similarities and differences \lenstro and \glee lens models \label{app:ssec:model_comp}}

\begin{table}[h]
    \caption{Overview of similarities and differences between \lenstro and \glee lens models.}
    \label{app:tab:lens_model_comp_list}
    \renewcommand{\arraystretch}{1.2}
    \centering
    \begin{tabular}{l c}
        \hline\hline
        Modeling aspect & Shared\tablefootmark{\dag} \\
        \hline
        HST (three filters) + time delays constraints    & \cmark \\
        Main deflector mass parametrization      & \cmark \\
        External shear component                 & \cmark \\
        Perturbers mass profiles                 & \cmark \\
        PSF subsampling                          & (\cmark) \\
        Main deflector light  parametrization    & (\cmark) \\
        Pixel selection (imaging likelihood)     & \xmark \\
        Overall pixel uncertainties (noise map)               & \xmark \\
        Local boosting of pixel uncertainties               & \xmark \\
        PSF model (initial \& corrections) \tablefootmark{\ddag}       & \xmark \\
        Ray-tracing subsampling                  & \xmark \\
        Extended source source light \tablefootmark{\S} & \xmark \\
        Priors on model parameters          & \xmark \\
        \hline
    \end{tabular}
    \tablefoot{
        \tablefoottext{\dag}{$\checkmark$ indicates the two codes share the same
            approach while $\times$ indicates they differ; ($\checkmark$) indicate aspects that were not agreed upon by the two teams, but that coincidentally became partly comparable (i.e., Sérsic profiles for the lens light, subsampling of the PSF in \ir).}
        \tablefoottext{\ddag} We performed \glee lens models both with their own initial and corrected PSF as well as the corrected PSF from \lenstro.
        \tablefoottext{\S}{Pixelated source in \glee, analytical (Sérsic+shapelets) in \lenstro.}
    }
\end{table}

To better grasp the similarities and differences between the \lenstro and \glee lens models, we compare them side-by-side in Table Table~\ref{app:tab:lens_model_comp_list}. The main differences lie in the procedure to reconstruct the quasar host, the PSF initial models and iterative corrections, as well as pixel-level uncertainty boosting. In addition, the two codes do not model the uncertainty per pixel in the same way. The background noise component has been estimated independently between the two teams, and the shot noise component is estimated based on the model flux in \lenstro while it remains fixed to the observed flux in \glee.

\begin{figure}[!th]
    \centering
    \includegraphics[width=\linewidth]{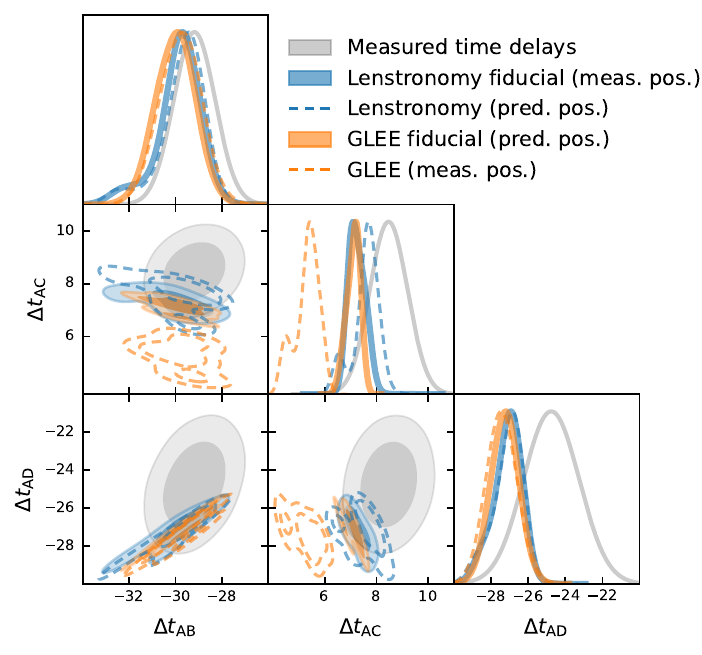}
    \caption{Distributions of time delays from \lenstro and \glee lens models, compared to the measurements (taken from Table~\ref{tab:td_j1537}). Beyond the measurements in gray, the colored solid lines and filled contours show the fiducial predictions used within each code to evaluate the time-delay likelihood during lens modeling. The fiducial choice for \lenstro is to evaluate the time delays at the measured image positions, while it is at the predicted image positions for \glee. The dashed lines and contours show the alternative choice, for reference only as they are not used in the analysis.}
    \label{app:fig:model_time_delays}
\end{figure}

Both codes included the measured time delays as constraints, jointly fitted with the imaging data. We show in Fig.~\ref{app:fig:model_time_delays} the posterior distributions of the time delays from each code. There are generally two possibilities to evaluate the time delays, either at the measured image positions (i.e., the position fitted to the pixels) or at the predicted image positions (i.e., the positions obtained by solving the lens equation). We show both in Fig.~\ref{app:fig:model_time_delays} for completeness; however, \lenstro uses the measured positions while \glee uses the predicted positions during modeling. Overall the time delays are well fitted by the two codes. The longest delay AB is the better fitted, while the AC and AD posterior delays are in slight disagreement ($\sim 1\sigma$) with the measurements.

\section{Dynamical modeling details}

\subsection{Surface brightness profile for dynamical models \label{app:ssec:lens_light_model}}

\begin{figure*}[!th]
    \centering
    
    \includegraphics[width=0.8\linewidth]{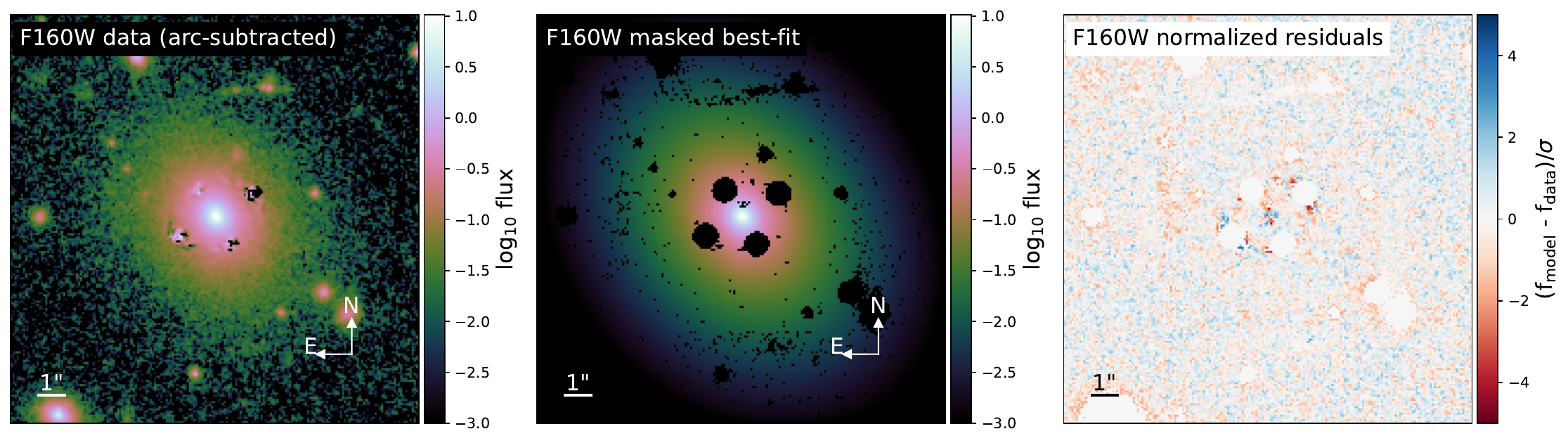}
    \includegraphics[width=0.8\linewidth]{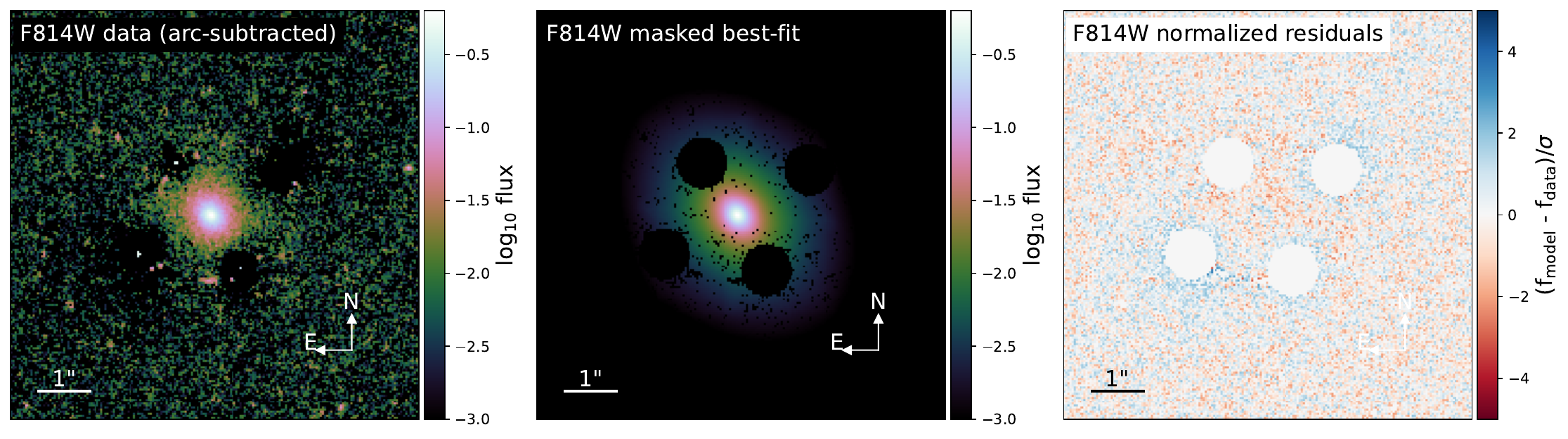}
    \caption{Surface brightness model fitting for dynamical modeling for HST filters \ir and \uvisone. The double elliptical Sérsic profile is fitted on large cutouts compared to those used for lens modeling, in particular to contain the full extent of the galaxy in the \ir filter. The resulting profiles are then used to construct a stitched model shown in Fig.~\ref{app:fig:stitched_lens_light}.}
    \label{app:fig:lens_light_fitting}
\end{figure*}

\begin{figure}[!th]
    \centering
    \includegraphics[width=\linewidth]{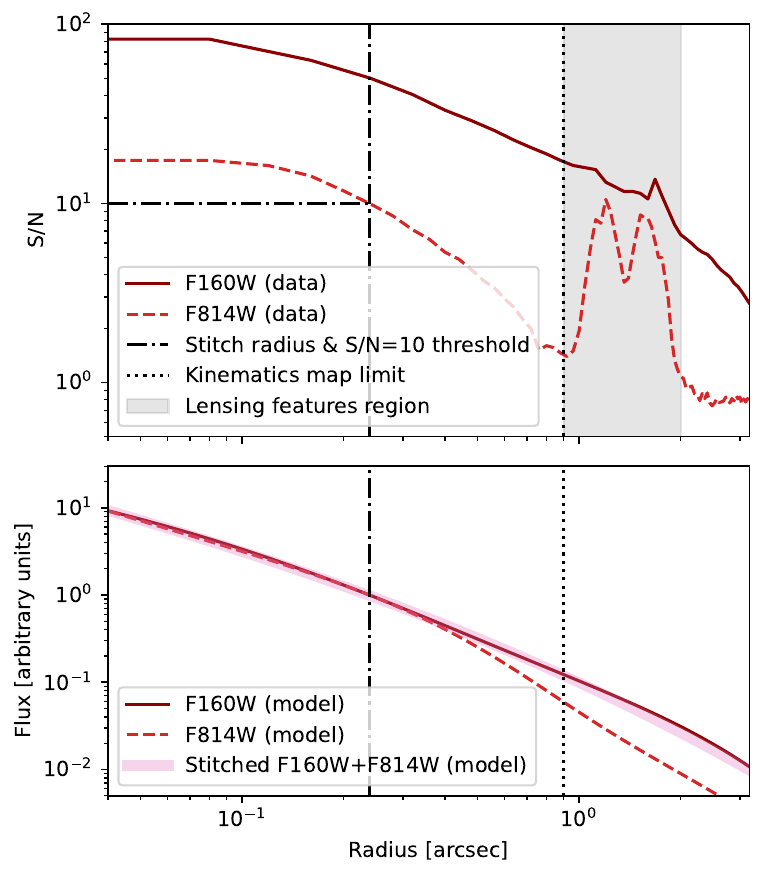}
    \caption{Surface brightness model of the main deflector used in our dynamical models. The top panel shows the observed radial profiles in two HST filters, and how we find a ``stitch radius'' $r_{\rm stitch}=0\farcs24$ at which we transition from \uvisone ($0\farcs04$/pixel) to \ir ($0\farcs08$/pixel). The bottom panel shows double-Sérsic models to each filter separately fitted to the images (from Fig.~\ref{app:fig:lens_light_fitting}), and the resulting double-Sérsic model fitted to the stitched profile.}
    \label{app:fig:stitched_lens_light}
\end{figure}

For predicting stellar kinematics maps of the main deflector, we need a model of its surface brightness. This model needs to be axisymmetric and deprojected using MGE in order to be used in \textsc{JamPy}. Our lens models shown in Fig.~\ref{fig:bestfit_lens_models} include such a component. However, since our lens modeling cutouts do not encompass the full extent of the galaxy in the \ir filter, we performed new fits on larger cutouts, still assuming a double elliptical Sérsic profile (Eq.~\ref{eq:def_sersic_ellipse}), with joint centroid, axis ratio and position angle. We fit each filter separately and join the axis ratios and position of the two Sérsic profiles in a given filter. The resulting fits are shown in Fig.~\ref{app:fig:lens_light_fitting}.

We then use these two-dimensional models in \ir and \uvisone and stitch them radially, using a procedure similar to \citet{2026A&A...707A.314S}. We require a minimum S/N of 10 (azimuthally averaged) in the \uvisone filter before transitioning to the \ir filter, a condition that is reached at radius $r_{\rm stitch}=0\farcs24$. The axis ratio $q_\ell$ and position angle $\phi_\ell$ being almost identical in the two filters, we take their values from the \ir fit because it is constrained by significantly more pixels compared to \uvisone. Our dynamical model from Sect.~\ref{sec:dynamical_weighting} use MGE decomposition of the stitched radial profile, with ellipticity and orientation information encoded in $q_\ell$ and $\phi_\ell$ to obtain the two-dimensional surface brightness $\Sigma_\ell$. The effective (half-light) radius corresponding to the stitched profile is $\reff = 1\farcs57 \pm 0\farcs05$.

\subsection{Radial stellar kinematics profiles}

\begin{figure}[!ht]
    \centering
    \includegraphics[width=\linewidth]{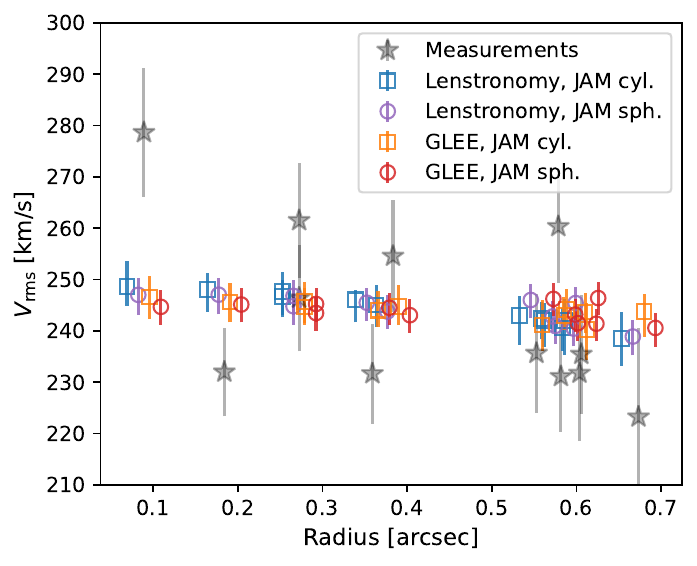}
    \caption{Best-weight dynamical models for each lens models (\lenstro and \glee) and each choices of JAM parametrization (cylindrically or spherically aligned stellar orbits), visualized as a radial profile rather than the fitted maps (Fig.~\ref{fig:dynamical_model_posterior}). Each model value is slightly shifted horizontally to better discriminate between the models.}
    \label{app:fig:dynamical_model_profiles}
\end{figure}

While our JAM model predicts stellar kinematics in two dimensions (right panel of Fig.~\ref{fig:dynamical_model_posterior}), it is informative to visualize the fit along the radial dimension. Figure~\ref{app:fig:dynamical_model_profiles} shows the $V_{\rm rms}$ in each bin centroid, for both the measurements and the four models. The larger mass density slope of the \lenstro lens model is visible, which provides a slightly better fit to the central bin with higher velocity dispersion.

\subsection{Full parameter space for dynamical modeling \label{app:ssec:dynamics_full_param}}

\begin{figure*}[!ht]
    \centering
    \includegraphics[width=\linewidth]{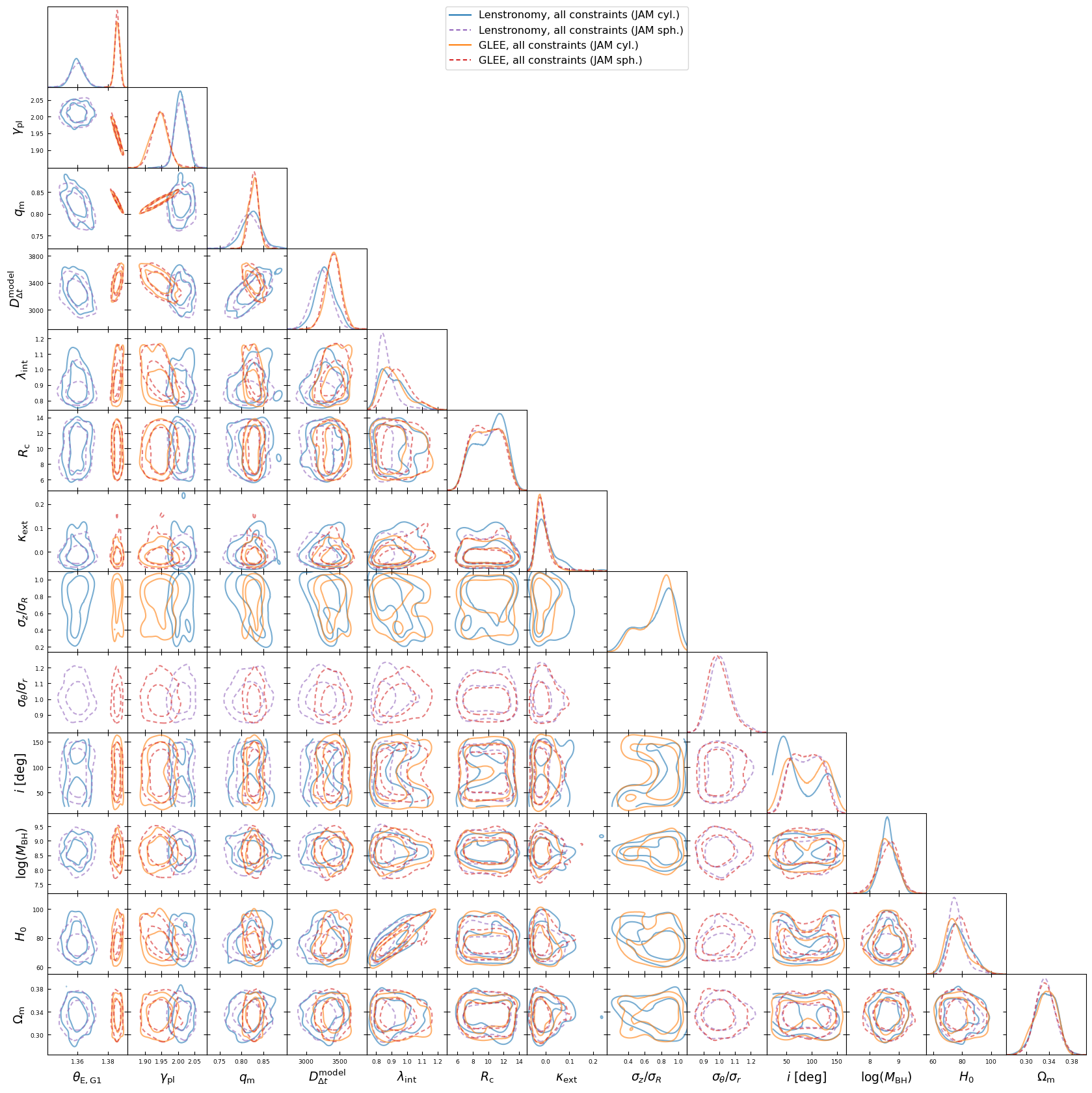}
    \caption{Full parameter space of the dynamical modeling step. From left to right, the first four parameters are primarily constrained by lensing (imaging and time delays), while the remaining of the parameter space is composed of dynamical, LoS and cosmological parameters. See Sect.~\ref{sec:dynamical_weighting} for the definition of all parameters and their prior distributions.}
    \label{app:fig:dynamical_model_full_posterior}
\end{figure*}

In Fig.~\ref{fig:dynamical_model_posterior} we show constraints on key parameters after including stellar kinematics and LoS information. However, as detailed in Sect.~\ref{sec:dynamical_weighting}, dynamical modeling with JAM and the MST parametrization requires slightly more parameters. For completeness we show in Fig.~\ref{app:fig:dynamical_model_full_posterior} the full set of parameters added during the dynamical modeling step, including the inclination angle, the black hole mass, and MST-approximating core radius. For visualizing the covariances with parameters constrained by strong lensing, the figure shows the G1 Einstein radius, mass density slope, mass axis ratio and (uncorrected) time-delay distance.

\section{Mass and light profile definitions \label{app:sec:profile_definitions}}

\subsection{Surface brightness}

In this paper we denote the surface brightness with $\Sigma_\ell$. Our fiducial model to describe $\Sigma_\ell$ is the elliptical version of the Sérsic profile \citep{Sersic1963}, defined as
\begin{align}
\label{eq:def_sersic_ellipse}
    \Sigma_{\ell,\rm sersic}(x,\,y) = I_{\rm eff}\,\mathrm{exp}\left[ -b_n \left\{ \left( \frac{\sqrt{q_\ell x^2 + y^2/q_\ell}}{\reff}\right)^{1/n_{\rm s}} - 1 \right\} \right]\ , 
\end{align}
where $I_{\rm eff}$ is the amplitude of the profile at the half-light radius $\reff$, $q_\ell$ is the axis ratio, $n_{\rm s}$ is Sérsic index. The term $b_n = 1.999\,n_{\rm s} - 0.3271$ ensures that $\reff$ corresponds to the half-light radius. The coordinate system is defined such that the $x$ and $y$ coordinates are along the major and minor axis respectively, after rotation by the position angle $\phi_\ell$. We note that throughout the paper, we use $\reff$ more generally to define the effective radius of the surface brightness distribution regardless if it was obtained using a single Sérsic profile (as in Eq.~\ref{eq:def_sersic_ellipse}), or numerically computed from a combination of Sérsic profiles or other profiles.

The chameleon profile \citep{Dutton2011} provides a different parametrization to but approximates well the Sérsic profile over the radial range relevant to galaxy-scale strong lensing. This profile is defined as
\begin{align}
\label{eq:def_chameleon}
\nonumber
    \Sigma_{\ell, \rm cham}(x, y) = \frac{I_0}{1 + q_\ell} 
    &\Bigg[ \frac{1}{\sqrt{x^2+y^2/q_\ell + 4w_{\rm c}^2/(1+q_\ell^2)}} \Bigg. \\
    &\Bigg. - \frac{1}{\sqrt{x^2+y^2/q_\ell + 4w_{\rm t}^2/(1+q_\ell^2)}} \Bigg] \ ,
\end{align}
where $I_0$ is a normalization factor, $w_{\rm c}$ and $w_{\rm t}$ can be interpreted as core and truncation radii, respectively. Elliptical coordinates parameters $q_\ell$ and $\phi_\ell$ remain the axis ratio and position angle as in Eq.~\ref{eq:def_sersic_ellipse}, respectively. 

\subsection{Mass density and shear \label{app:ssec:mass_profiles}}

We describe the (total) mass density of the main deflector via its convergence $\kappa\equiv\Sigma_{\rm m}/\Sigma_{\rm crit}$, where $\Sigma_{\rm m}$ is the is the physical mass density (projected onto deflector's redshift plane) and $\Sigma_{\rm crit}$ is the critical mass density (Eq.~\ref{eq:sigma_crit_ddt_dd}). For the lens modeling step (Sect.~\ref{sec:lens_modeling}), we assume the convergence follows an elliptical power-law (EPL) profile defined as \citep{2015A&A...580A..79T}
\begin{align}
\label{eq:def_epl}
    \kappa_{\mathrm{EPL}}(x, y) &= \frac{3 - \gamma_{\rm pl}}{2} \left[ \frac{\thetaE}{\sqrt{q_\mathrm{m}x^2 + y^2/q_\mathrm{m}}}\right]^{\gamma_{\rm pl} - 1}\ ,
\end{align}
where $\gamma_{\rm pl}$ is the logarithmic power-law slope, $q_{\rm m}$ is the axis ratio and \thetaE is the Einstein radius (more precisely, the circularized Einstein radius\footnote{See for example the definition at \url{https://github.com/lenstronomy/lenstronomy/blob/main/lenstronomy/LensModel/Profiles/epl.py}}. The coordinate system is defined such that the $x$ and $y$ coordinates are along the major and minor axis respectively, after rotation by the position angle $\phi_{\rm m}$. A limiting case of the EPL profile is the singular isothermal sphere (SIS), which has a fixed isothermal slope ($\gamma=2$) and no ellipticity ($q_{\rm m}=1$).

To account potential deviations to the power-law mass model, we introduce an additional uniform shear field. The shear is parameterized by a distortion amplitude $\gamma_{\rm ext}$ and orientation $\phi_{\rm ext}$. The lensing potential produced by this shear component is, in polar coordinates:
\begin{align}
    \label{eq:def_shear}
    \psi_{\rm ext}(r) = \frac12\,\gamma_{\rm ext}\,r^2\cos\left(\phi - \phi_{\rm ext}\right) \ ,
\end{align}
where $r$ and $\phi$ are standard polar coordinates.

We note that the zero-point for position angles (for elliptical profiles and shear components) typically differ between modeling codes. In \glee, angles are measured counterclockwise from the positive x axis while it is clockwise from positive x axis in \lenstro (the x-axis can furthermore be oriented differently in each code, along image axes or sky RA-Dec coordinates). In this work, the correspondance between the angles returned by \glee and \lenstro is $\phi_{\glee} = 90^\circ -\phi_{\lenstro}$.

\end{appendix}

\end{document}